\documentclass[journal]{IEEEtran}
\IEEEoverridecommandlockouts
\usepackage[T1]{fontenc}

\usepackage{cite}
\usepackage{amsmath,amssymb,amsfonts,amsthm}
\usepackage{array}
\usepackage{booktabs}
\usepackage{colortbl}
\usepackage{graphicx}
\usepackage{blindtext}
\usepackage{textcomp}
\usepackage[dvipsnames]{xcolor}
\usepackage{braket}
\usepackage{subcaption}
\usepackage{float}
\usepackage{tikz}
\usepackage{quantikz}
\usepackage[linesnumbered,ruled,noend]{algorithm2e}

\usepackage{pgfplots}
\usepgfplotslibrary{groupplots}
\pgfplotsset{compat=1.18}

\usetikzlibrary{patterns}
\usetikzlibrary{spy, calc}

\usepackage{hyperref}
\hypersetup{
    colorlinks = true,
    linkcolor = BlueViolet,
    citecolor = BlueViolet,
    filecolor = BlueViolet,
    urlcolor = BlueViolet,
}

\newtheorem{proposition}{Proposition}
\newtheorem{theorem}{Theorem}
\newtheorem{corollary}{Corollary}
\newtheorem{definition}{Definition}
\theoremstyle{plain}
\theoremstyle{remark}
\newtheorem*{remark}{Remark}

\newtheorem*{resolution}{Whatever Channel Resolution}

\newtheorem*{configspace}{Full Configuration Space}

\newcommand{\RN}[1]{%
    \textup{\uppercase\expandafter{\romannumeral#1}}%
}

\def\BibTeX{{\rm B\kern-.05em{\sc i\kern-.025em b}\kern-.08em
    T\kern-.1667em\lower.7ex\hbox{E}\kern-.125emX}}

\begin{document}

\title{A Resource-Driven Framework for Configurable Entanglement in Quantum Networks}

\author{
    \IEEEauthorblockN{
    Francesco Mazza,~\IEEEmembership{Graduate~Student~Member,~IEEE}, 
    Claudio Pellitteri, \\Angela~Sara~Cacciapuoti,~\IEEEmembership{Senior~Member,~IEEE}, and Marcello~Caleffi,~\IEEEmembership{Senior~Member,~IEEE}
    \thanks{The authors are with the \href{www.quantuminternet.it}{www.QuantumInternet.it} research group, University of Naples Federico II, Naples, 80125 Italy.}
    \thanks{Corresponding author: Angela Sara Cacciapuoti, angelasara.cacciapuoti@unina.it}
    \thanks{This work has been funded by the European Union under Horizon Europe ERC-CoG grant QNattyNet, n.101169850. Views and opinions expressed are however those of the author(s) only and do not necessarily reflect those of the European Union or the European Research Council Executive Agency. Neither the European Union nor the granting authority can be held responsible for them.}}
}  

\maketitle

\begin{abstract}
Shared multipartite entanglement defines a ``\textit{whatever channel}'', i.e., a latent communication substrate that does not determine \textit{a priori} which end-to-end entangled links are activated, but can be configured to support different entanglement-connectivity graphs through Local Operations and Classical Communication (LOCC). Building on this, we propose a resource-driven framework in which multipartite entanglement is treated as a \textit{programmable resource} that induces a space of admissible entanglement-graph configurations. Within this framework, connectivity provisioning emerges as a particular instance of a more general resource reconfiguration process. To support this paradigm, we introduce a set of structural design parameters that characterize the operational degrees of freedom of the resource and define the admissible transformations independently of the specific mechanism used to realize them. We then formalize Entanglement Rolling as a measurement-based protocol that operates over the induced configuration space, enabling the systematic reconfiguration of the shared resource across a family of multipartite states. Finally, we analyze the proposed framework under relevant noise processes. Leveraging the Noisy Stabilizer Formalism (NSF), we derive closed-form noise maps that characterize the effect of noise on the resource transformations and show that the instantiated resources remain distillable over a broad range of noise parameters.
\end{abstract}


\section{Introduction}
\label{sec:01}
The Quantum Internet~\cite{CacCal-26, CalCac-25, CacCalIll-26, Kim-08, DurLamHeu-17, CacCalVan-20, AzuEcoElk-23, PirDur-19} is expected to enable applications beyond the capabilities of classical networks, such as distributed quantum computing~\cite{CirEkeHue-99, HayMor-15, CalAmoFer-22}, unconditionally secure communications~\cite{GisRibTit-02, PirAndBan-20}, and enhanced quantum sensing~\cite{GioLloMac-11, KesLovSus-14, SekWolDur-20, GiaWinCon-25}. At the core of this vision lies \textit{quantum entanglement}, the fundamental resource of the network~\cite{JozLin-03, AviRozWeh-23, IneVarSca-23, ChuRamAni-24, AbaCubMai-25}. Bipartite entanglement enables point-to-point primitives such as teleportation~\cite{CirZolKim-97, CacCalVan-20}, while multipartite entanglement defines more general and complex communication structures~\cite{CacIllCal-23, RamPirDur-21, IllCalVis-23}, supporting flexible and on-demand connectivity patterns~\cite{IllCalMan-22, CheCacCal-26, MazCalCac-25} beyond the constraints of the underlying physical network graph. Due to this flexibility, multipartite entanglement has been increasingly regarded as a versatile \emph{network resource}. 

Nevertheless, the existing literature has predominantly treated multipartite entanglement  as a tool for connectivity provisioning, focusing on how to distribute resources and react to sets of communication requests, as detailed in Sec.~\ref{sec:01-A}. In this perspective, entanglement manipulation is typically designed to satisfy specific connectivity demands, e.g., by extracting Bell pairs or establishing end-to-end entanglement between selected nodes.

In contrast, this work adopts a different viewpoint. Rather than focusing on how to satisfy given requests, we investigate how shared multipartite entanglement can be treated as a \textit{programmable resource} that enables the systematic instantiation of entanglement-based functionalities.

To this end, we introduce the concept of \textit{whatever}\footnote{The authors gratefully acknowledge John D. Day for suggesting this name.} channel\footnote{Here, the term \textit{channel} is not used in the conventional sense of a physical medium underlying a communication link. Rather, it denotes an entanglement-defined communication structure, in which the involved nodes may be physically distant, while still being neighbors in the entanglement connectivity graph \cite{IllCalMan-22}.}, namely, a shared multipartite entangled resource that does not determine \textit{a priori} which end-to-end entangled links are instantiated, but supports the instantiation of different subsets of links through suitable operations on the shared state.
In this sense, the \textit{whatever} channel is a latent communication substrate that can be configured to realize different entanglement-connectivity patterns depending on how it is manipulated.

Building on this programmable nature of the \textit{whatever} channel, we formalize a \emph{resource-driven framework}, where different entanglement-based functionalities can be realized by appropriately transforming the underlying resource. 
More precisely, the \textit{whatever} channel induces a well-defined space of admissible entanglement-graph configurations, rather than a fixed pattern. Accordingly, entanglement manipulation is not viewed as a one-shot extraction process, but as a structured transformation that operates over this configuration space. 

To support this process, we introduce (in Sec.~\ref{sec:03}) a set of structural definitions that formalize the operational degrees of freedom of the resource. These constructs provide a functional characterization, rather than a merely descriptive one, as they directly determine the admissible transformations of the resource and, consequently, the set of entanglement-graph configurations that can be instantiated. In particular, the proposed framework is not tied to a specific resource instance, but applies to a family of multipartite resource states, characterized by structural design parameters. These parameters induce the admissible transformations of the resource and define the corresponding space of achievable entanglement-graph configurations, independently of the mechanism used to realize them.

Within this framework, we exploit and mathematically formalize Entanglement Rolling, first introduced in~\cite{MazCalCac-25} for proximity reduction within the same resource model, as a measurement-based protocol that enables the systematic reconfiguration of the shared resource across the considered family of states. Rather than acting as a mechanism for extracting specific entangled pairs, Entanglement Rolling operates over the induced configuration space, enabling the systematic exploration of the configurations it reaches from these states as a general realization mechanism. In this sense, it applies uniformly across the considered family of resource states. The Entanglement Rolling relies on a hierarchical organization of network nodes, potentially supported by control-plane coordination, and it is therefore consistent with emerging quantum network architectures~\cite{CalCac-25,CacCal-26}.

Since realistic quantum networks operate under decoherence, we also analyze the proposed framework under noisy conditions. By leveraging the Noisy Stabilizer Formalism \cite{RuiDur-23,AigRuiDur-25}, we derive closed-form noise maps that characterize the effect of noise on the resource transformations. Furthermore, we demonstrate that the instantiated resources remain distillable over a broad range of depolarizing and memory dephasing noise.

Our contributions can be summarized as follows:
\begin{itemize}
    \item[i.] we introduce the concept of \textit{whatever channel} as a shared multipartite resource and propose a \textit{resource-driven framework} for its programmable resolution over a family of resource states;
    \item[ii.] we formalize Entanglement Rolling, characterizing each of its measurement steps in closed form, as a measurement-based reconfiguration protocol that operates over the induced configuration space, enabling controlled transitions among admissible entanglement-graph configurations; as a key instance, we prove that, followed by a local isolation step at Tier1, it achieves the maximum number of concurrently instantiable Bell pairs over the considered resource family, with the minimum number of measurements in all the explored instances;
    \item[iii.] we derive closed-form noise maps for the resource transformations induced by Entanglement Rolling and by the isolation step at Tier1, together with the fidelity of the instantiated Bell pairs, providing an analytical characterization of the framework under relevant noise processes;
    \item[iv.] we evaluate the performance of the proposed framework under both depolarizing and time-dependent dephasing noise, characterizing the attainable resource fidelities and operating regimes within the considered noise models.
\end{itemize}

The companion implementation checks the analytical results and reproduces the numerical experiments~\cite{repo}.

The remainder of this paper is organized as follows. A summary of the related works is provided in Sec.~\ref{sec:01-A}. Sec.~\ref{sec:02} provides the necessary background on multipartite entangled states and their noisy manipulation. In Sec.~\ref{sec:03}, we introduce the system model and the communication paradigm of interest. Sec.~\ref{sec:04} presents the Entanglement Rolling protocol and its properties. Sec.~\ref{sec:05} describes  the systematic reconfiguration of the resource state through Entanglement Rolling and in Sec.~\ref{sec:06} we analyze the effect of noisy entanglement manipulation. In Sec.~\ref{sec:07} we conclude the paper.

\begin{figure}
    \centering
    \includegraphics[width=\linewidth]{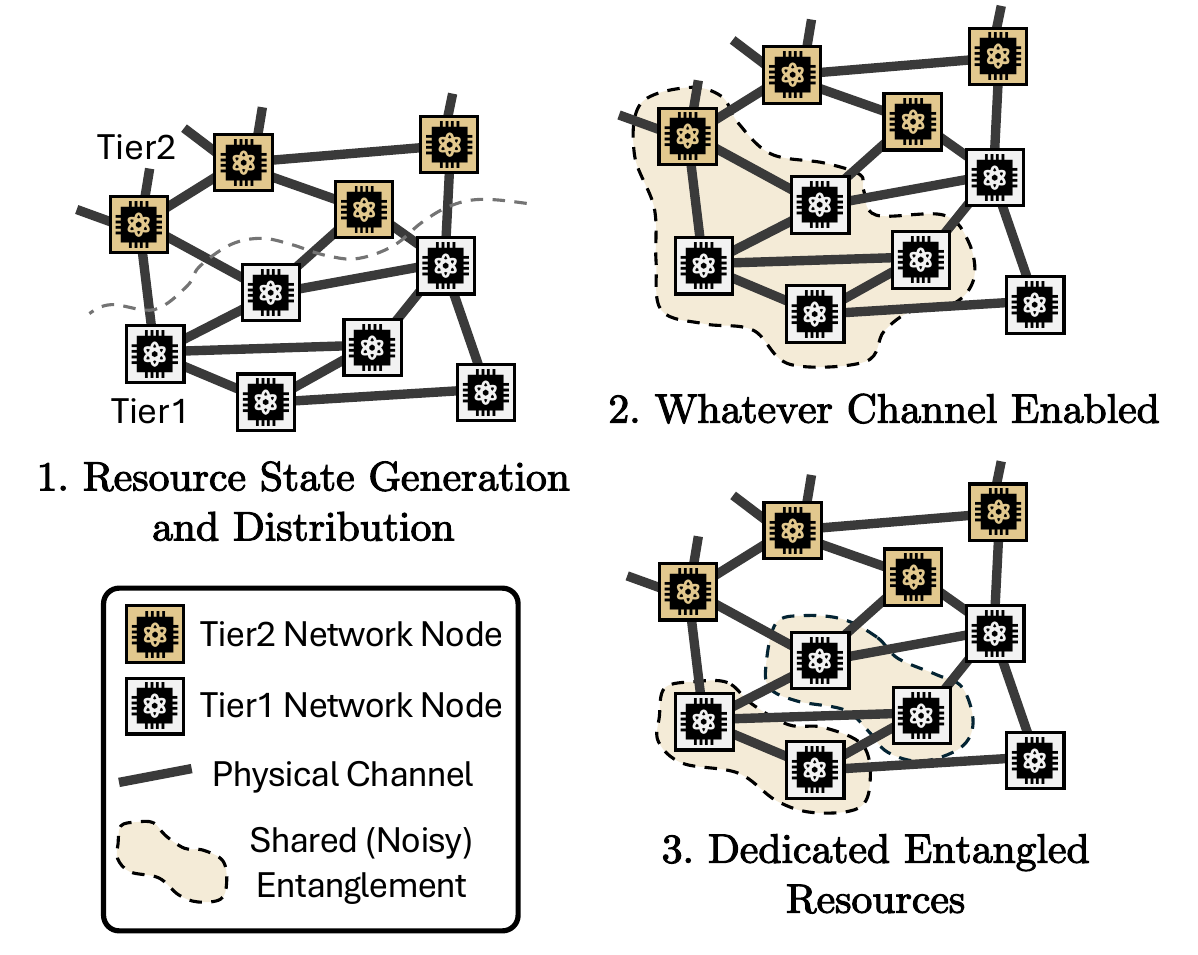}
    \caption{High level representation of our research problem. The generation and distribution of a multipartite resource state (1) establishes a \textit{whatever channel} between a subset of network nodes (2). By applying LOCC operations on the shared resource state, entanglement can be reconfigured and shared between a subset of network nodes (3), for instance, parallel communication resources can be extracted through the resolution of the \textit{whatever channel}.}
    \label{fig:01}
\end{figure}

\subsection{Related work}
\label{sec:01-A}
Existing literature on multipartite entangled resources can be broadly grouped along two main directions: i) generation and distribution; ii) utilization for connectivity provisioning. Regarding the first direction, a substantial body of literature addresses how multipartite resource states can be generated and distributed over quantum networks, with performance largely dominated by technological constraints and network scale. Representative approaches include graph-state generation via quantum emitters~\cite{ButBarEco-17, RusBarEco-19, ThoRusMor-22} and fusion operations~\cite{ThoRusRem-24, MenFauCha-25}, as well as network-level distribution mechanisms encompassing purification, and repeater-based schemes~\cite{AzuEcoElk-23, AzuTamLo-15, RuiRamWal-26}. %

Regarding instead the second direction, multipartite resources have been extensively studied as \textit{connectivity-provisioning} tools, where local operations reshape the resource to satisfy point-to-point entanglement requests~\cite{DahWeh-18, DahHelWeh-20}. In this context, graph and cluster states~\cite{HeiEisBri-04, HeiDurEis-06} have emerged as the predominant resource models, as they provide a convenient representation that enables structured manipulation and analytical tractability. This perspective underlies a wide range of works on Bell-pair routing~\cite{HahPapEis-19}, quantum repeater architectures~\cite{AzuTamLo-15, AzuEcoElk-23, RuiRamWal-26}, and resource allocation across network scales~\cite{MazCalCac-25, RamPirDur-21, CalCac-25, MazRamIll-25, BhaGoo-26, RamIllMaz-26}. 

The present work departs from the above perspectives in a fundamental way: rather than treating graph-state manipulation as a means to satisfy predefined connectivity requests, we consider multipartite entanglement as a programmable resource that induces a space of admissible configurations. In this view, connectivity provisioning becomes one possible outcome of a more general resource reconfiguration process. Accordingly, our goal is to define a resource-driven framework that characterizes how a shared multipartite state can be systematically reconfigured. Within this framework, mechanisms such as Entanglement Rolling provide a concrete way to operate over the induced configuration space.

Given the context and the motivations above, our work has been inspired by~\cite{MazCalCac-25}, which manipulates GTL states within a quantum local area network to engineer specific entanglement graphs, and introduces Entanglement Rolling to reduce the proximity between two given clients. We follow parts of the terminology and the notation used within this preliminary work and adopt and extend its GTL states, structural parameters and proximity metric, while addressing a different problem: how Entanglement Rolling navigates the configuration space induced by the shared GTL family of states, how the reached configurations resolve into disjoint resources and how this relates to a hierarchical and noisy setting. To this end, we generalize Entanglement Rolling to any choice of the support vertices, which makes the proximity reduction of~\cite{MazCalCac-25} a special case of a more general manipulation stage. We introduce a two-stage resolution that instantiates the maximum number of concurrent Bell pairs for every GTL state, following the intuition behind the previous work. Indeed, the results obtained by~\cite{MazCalCac-25} are naturally contained in our framework, which allows for arbitrary Pauli measurements. Finally, by building on existing results on noisy graph-state manipulation~\cite{RuiDur-23} and recent quantum network architectures~\cite{CalCac-25, MazCalCac-25}, we provide a noise-aware and operational treatment that connects resource-state structure with practical deployability.

\section{Preliminaries}
\label{sec:02}
Here, we introduce graph states and the Noisy Stabilizer Formalism (NSF), the two mathematical tools used throughout the paper to model and manipulate noisy resources.

A graph state $\ket{G}$ is associated to a graph $G = (V,E)$, where $V$ is the set of vertices and $E$ is the set of edges. To each vertex of the graph corresponds a qubit, and the entanglement between different qubits is represented by the application of a controlled-$Z$ gate, thus encoding the edges of the graph. 
The formalism of graph states allows for a straightforward description of entanglement relations between subsystems as well as a natural way of describing their manipulations, as detailed in \cite{HeiEisBri-04, HeiDurEis-06}. Interestingly, graph states are stabilizer states: $\ket{G}$ is the unique common $+1$ eigenstate of an associated stabilizer group.

\renewcommand{\arraystretch}{1.8}
\begin{table}[t]
    \centering

    \fontsize{9pt}{9pt}\selectfont
    \colorlet{headercolor}{BlueViolet!20}
    \colorlet{rowcolor}{BlueViolet!5}
    \begin{tabular}{c p{0.62\linewidth}}
        \toprule
        \rowcolor{headercolor}
        \multicolumn{1}{p{0.27\linewidth}}{\textbf{Operator $\mathbf{O}$ on qubit $a$}} & \multicolumn{1}{c}{\textbf{Updated $j$-th Noise operator}} \\ 
        \midrule

        \cellcolor{rowcolor}$\mathbf{M}_{z,\pm}^{(a)}$ & \cellcolor{rowcolor}$\displaystyle\mathbf{\tilde{\Lambda}}_{j,\alpha\beta} = 
            \begin{cases}
                \prod_{k \in N_a} Z_k^{\beta} & \text{if $j = a$}\\
                
                Z_j^{\alpha}\prod_{k \in N'_j} Z_k^{\beta} & \text{otherwise}
            \end{cases}$ \\ 
        \midrule

        $\mathbf{M}_{y,\pm}^{(a)}$  & $\displaystyle \mathbf{\tilde{\Lambda}}_{j,\alpha\beta} = 
            \begin{cases}
                \prod_{k \in N_a} Z_k^{\alpha +\beta} & \text{if $j = a$} \\
                
                Z_j^{\alpha + \beta}\prod_{k \in N'_j} Z_k^{\beta} & \text{if $j\in N_a$} \\
                
                Z_j^{\alpha}\prod_{k \in N'_j} Z_k^{\beta} & \text{otherwise}
            \end{cases}$ \\
        \midrule

        \cellcolor{rowcolor}$\mathbf{M}_{x,\pm}^{(a)}$  & \cellcolor{rowcolor}$\displaystyle \mathbf{\tilde{\Lambda}}_{j,\alpha\beta} = 
            \begin{cases}
                Z_{b_0}^{\alpha}\prod_{k \in N_{b_0}} Z_k^{\alpha} & \text{if $j = a$} \\
                
                Z_{b_0}^{\beta}\prod_{k \in N'_{b_0}} Z_k^{\alpha} & \text{if $j = b_0$} \\
                
                Z_j^{\alpha} \prod_{k \in N'_j} Z_k^{\beta} & \text{otherwise}
            \end{cases}$ \\
        \bottomrule
            
    \end{tabular}
    \caption{Updating rules for Pauli noise operators corresponding to the application of operator $\mathbf{O}$ on qubit $a$. Here $N'_j$ denotes the neighborhood of qubit $j$ in the graph \emph{after} the application of $\mathbf{O}$. The exponents $\alpha,\beta\in\{0,1\}$ are binary indices with $Z^0=\mathbb{I}$ and $Z^1=Z$. Table reproduced from~\cite{RuiDur-23}.}
    \label{tab:nsf_update_rules}
\end{table}

\begin{definition}[Stabilizer Operators of Graph States]
    \label{def:stabilizer_operators}
    The stabilizer operators $\{K_a\}_{a \in V}$ of an $n$-qubit graph state $\ket{G}$, associated with the graph $G=(V,E)$, are the $n$ commuting Hermitian $n$-qubit Pauli operators satisfying $K_a \ket{G} = \ket{G}$, for all $a \in V$ with
    \begin{equation}
        \label{eq:stabilizer_operators}
        K_a = X_a \prod_{b \in N_a} Z_b,
    \end{equation}
    where $X_a$ and $Z_b$ denote the Pauli-X and Pauli-Z operators acting on qubits $a$ and $b$ respectively, and $N_a \subseteq V$ denotes 
    the neighborhood of vertex $a$ in $G$.
\end{definition}

\begin{remark}
    The symbol $U_j \in \{\mathbb{I}_j, X_j, Y_j,Z_j\}$ implicitly refers to the application of identity operators to all qubits except $j$:
\begin{equation}
    \label{eq:operator_conv}
    U_j = (\mathbb{I}^{(1)} \otimes \dots \otimes U^{(j)} \otimes \dots \otimes \mathbb{I}^{(n)}) .
\end{equation}
\end{remark}

The application of specific local unitaries and single-qubit Pauli measurements on a graph state yields, up to local corrections, another graph state. The corresponding transformation can be described directly at the level of the underlying graph via simple graph operations.  
These operations can be concisely described by a \textit{manipulation operator} $\mathbf{O}$ (e.g. a Pauli measurement on qubit (a) $\textrm{M}_\xi = \bra{\xi,\pm}^{(a)}\otimes (U_{\xi,\pm}^{(a)})^{\dagger} P_{\xi,\pm}^{(a)}$, with $\xi \in \{x,y,z\}$ or a local complementation unitary), applied to the graph state $\ket{G}$ to obtain a new graph state $\ket{G'}$.
We refer the readers to \cite{HeiDurEis-06, HeiEisBri-04} for a detailed discussion on the correspondence between local operations on graph states and their resulting associated graph.

The Noisy Stabilizer Formalism (NSF) is a powerful method to describe the manipulation of noisy graph states~\cite{RuiDur-23, AigRuiDur-25}. Specifically, instead of applying manipulation operators directly to a noisy graph state, NSF allows the application of manipulation operators to the underlying pure state $\varrho = \ket{G}\bra{G}$, while updating the noise maps acting on the state.
Thanks to the noisy stabilizer formalism, the noise operators $\mathbf{\Lambda}_{j,\alpha\beta}$ of each (Pauli) noise map $\mathcal{M}_j$ can be updated up to recurrent commutation rules, depending on the applied manipulation operator $\mathbf{O}_i$. Suppose we apply a set of manipulation operators $ \mathcal{O} = \{ \mathbf{O}_1,\dots,\mathbf{O}_k\}$ to the noisy graph state $\mathcal{M}_n \dots \mathcal{M}_1 \varrho$. 
The resulting state is given by: 
\begin{equation}
    \label{eq:noisy_stabilizer_formalism}
    (\mathbf{O}_k \dots \mathbf{O}_1)\mathcal{M}_n \dots \mathcal{M}_1 \varrho(\mathbf{O}_1^{\dagger} \dots \mathbf{O}_k^{\dagger}).    
\end{equation}
Thanks to the NSF, this can be equivalently written as:
\begin{equation}
    \label{eq:noisy_stabilizer_formalism_updated}
    \mathcal{\tilde M}_n \dots \mathcal{\tilde M}_1 (\mathbf{O}_k \dots \mathbf{O}_1)\varrho(\mathbf{O}_1^{\dagger} \dots \mathbf{O}_k^{\dagger}),
\end{equation}
where $(\mathbf{O}_k \dots \mathbf{O}_1)\varrho(\mathbf{O}_1^{\dagger} \dots \mathbf{O}_k^{\dagger}) = \varrho'$ is the noiseless manipulated graph state, according to manipulation operators $\mathbf{O}_i$ with $i \in \{1,\dots,k\}$, and $\mathcal{\tilde M}_j$ is the \textit{updated} noise map, with $j \in \{1,\dots,n\}$. 
Consequently, the resulting noisy graph given by the action of independent single-qubit noises is given by the application of multiple maps: $\mathcal{\tilde M}_n \dots \mathcal{\tilde M}_1 \varrho'$.
As a result, each updated noise map on a qubit $j$ is composed by updated noise operators $ \mathbf{\tilde{\Lambda}}_{j,\alpha \, \beta}$ whose update is completely described by recurrent commutation rules with respect to the applied operation~\cite{RuiDur-23}, and explicitly reported in Tab.~\ref{tab:nsf_update_rules} for Pauli measurements.

\section{System Model: Resource-Driven Entanglement Reconfiguration}
\label{sec:03}
We consider a quantum network in which a set of nodes share a multipartite entangled resource state. As discussed in Sec.~\ref{sec:01}, such a resource defines a \textit{whatever channel}, i.e., a programmable communication substrate that can be configured to realize different entanglement-connectivity patterns through LOCC. Indeed, the shared entanglement induces a space of admissible entanglement-graph configurations, determined by its structure and by the allowed set of local operations. Within this space, different configurations correspond to different realizable connectivity patterns among subsets of nodes. For graph states, which are in one-to-one correspondence with graphs, this space admits the following characterization.

\begin{configspace}
    Given a graph state $\ket{G}$, its full configuration space $\mathcal{G}$ is the set of the graph states obtained from $\ket{G}$ by single-qubit Clifford operations and Pauli measurements, i.e., its qubit-minors~\cite{DahHelWeh-20b, DahHelWeh-20}.
\end{configspace}

\begin{resolution}
    Given a whatever channel, its \textit{resolution} is the selection and instantiation of a specific entanglement-graph configuration from the admissible space, through LOCC applied to qubits stored at the network nodes.
\end{resolution}

Each resolution consumes one independently prepared resource state. A fresh preparation allows the network to select a different attainable configuration.

\begin{figure*}
    \centering
    \includegraphics[width=0.77\linewidth]{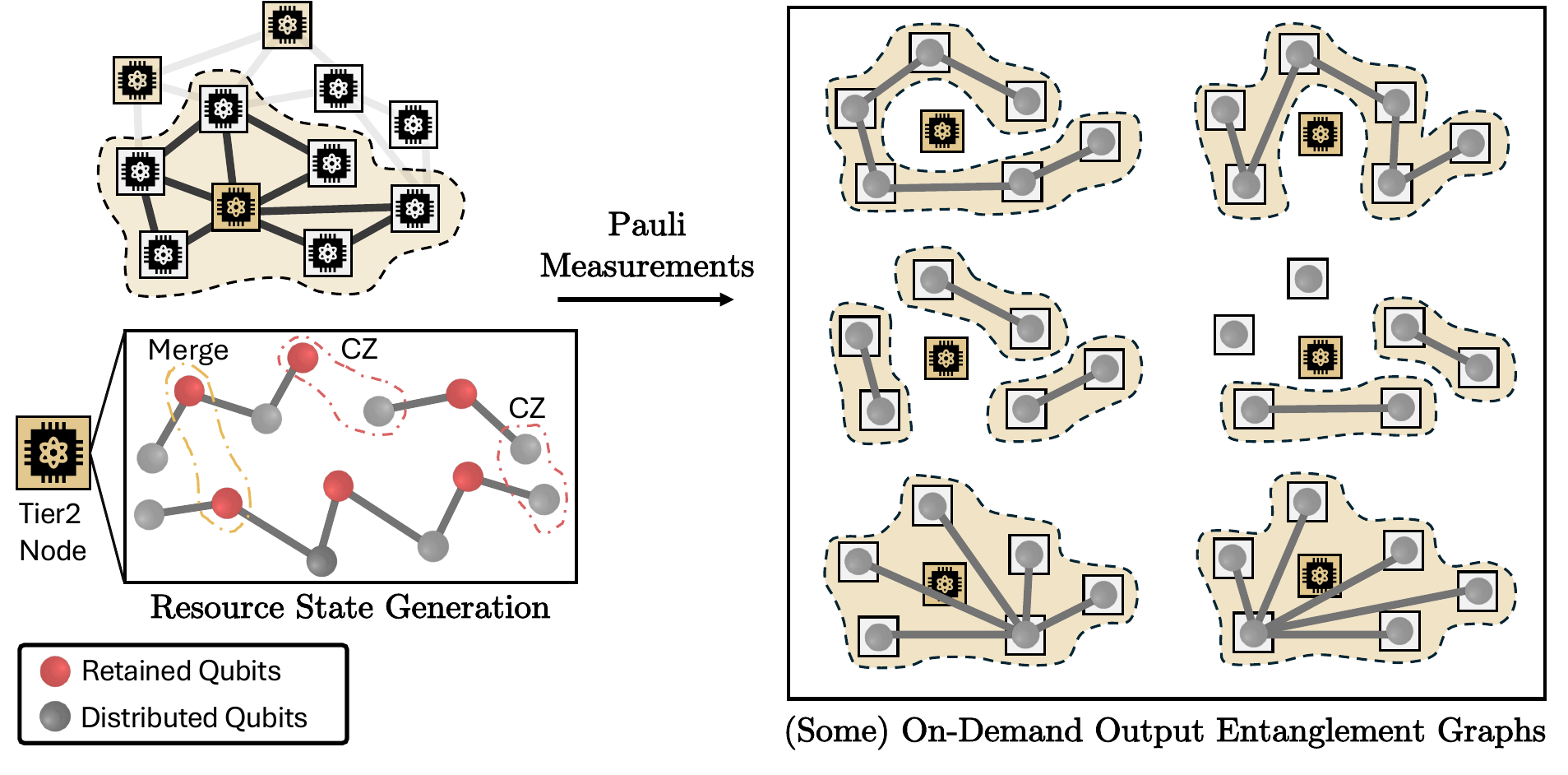}
    \caption{Pictorial representation of the system model and node hierarchy. Tier2 node(s) retains $n_o$ orchestration qubits $V_o$ of a shared two-colorable graph state, while the $\kappa$ peer qubits $V_c$ are distributed to Tier1 nodes. }
    \label{fig:02}
\end{figure*}

A key problem is therefore to characterize and systematically explore such a configuration space. In particular, we aim to identify: (i) the \textit{structural properties} of the resource state that determine the admissible transformations, and (ii) the \textit{mechanisms} that enable the controlled realization of useful configurations within that space.

Importantly, these two aspects are conceptually decoupled. The resource structure determines the constraints and the set of achievable configurations, independently of the realization mechanism. In contrast, concrete protocols operate over the induced space, by providing explicit procedures to transform the resource to satisfy specific tasks and network requests. In the following, we formalize this perspective by introducing a set of structural design parameters that characterize the resource and its admissible transformations, and by defining the operational model used to manipulate the shared entanglement.

\subsection{System Model and Communication Paradigm}
\label{sec:03-A}
Whatever channel resolution is not tied to a unique realization. In general, an arbitrary multipartite resource may induce an opaque space of admissible entanglement-connectivity configurations. For generic graph states, determining which configurations can be reached is intractable: even deciding whether Bell pairs can be obtained on a given set of qubits is NP-complete~\cite{DahHelWeh-20}. For this reason, we adopt a \textit{resource-engineering} perspective: rather than assuming an arbitrary shared state, we focus on \textit{parametrized} resource families, whose structure exposes controllable degrees of freedom. This yields an explicit characterization of the configurations reachable by acting on these degrees of freedom, enabling systematic control and providing a foundation to realize multiple entanglement-based functionalities via resource reconfiguration.

To this end, we adopt \textit{Generalized Tree-like} (GTL) graph states, a family of two-colorable graph states with proven networking applications \cite{MazCalCac-25, MazZhaChu-25}. GTL states are specified by structural design parameters that capture their topology and determine the admissible entanglement transformations they support. Their intrinsic bipartite structure induces a natural partition of qubits into two disjoint vertex sets, $V = V_o \cup V_c$, $V_o \cap V_c = \emptyset$, which we interpret operationally through a \textit{hierarchical network model}.

In fact, emerging Quantum Internet architectures distinguish network nodes in terms of computational resources and functional responsibilities~\cite{CalCac-25,CacCal-26}. In particular, \textit{Tier2 nodes} are high-capability devices that host and manipulate multipartite resources, whereas \textit{Tier1 nodes} are resource-constrained devices, that primarily apply local corrections and consume the provided entanglement. 

Within this model, Tier2 nodes collectively retain the qubits in $V_o$, termed \textit{orchestration qubits}, of a shared two-colorable graph state, while Tier1 nodes receive the distributed qubits in $V_c$, termed \textit{peer qubits}. 
Each Tier2 node can perform local Pauli measurements on its orchestration qubits independently, triggering the desired entanglement manipulation, regardless of how the remaining orchestration qubits are distributed among other Tier2 nodes. The resulting entanglement manipulation is coordinated across Tier2 nodes and delivered to Tier1 nodes through LOCC, yielding on-demand entanglement links among selected utilizers, as depicted in Fig.~\ref{fig:02}.

We note that the adopted \textit{hierarchical network model} is \textit{scale-adaptable}. Depending on the deployment, it can describe large-scale tiered architectures with multiple Tier2 nodes as in \cite{CalCac-25}, or compact settings in which a single node concentrates Tier2 functionality. Accordingly, all definitions, theorems, and results in this manuscript are independent of the number of Tier2 nodes. As a concrete example, a Quantum Local Area Network (QLAN)~\cite{MazCalCac-25, MazZhaChu-25} instantiates the same hierarchy at small scale, where a single orchestrator acts as the Tier2 node and the serving nodes act as Tier1 nodes.

\subsection{Resource State Design Parameters}
\label{sec:3.2}
To ensure practical realizability, we focus on two-colorable graph-state resources, which admit modular constructions from elementary one-dimensional (1D) cluster states via suitable merging operations\footnote{A merging operation between two vertices of different graph states combines them into a single vertex. In the optical graph-states literature this is often referred to as fusion.} \cite{ButBarEco-17, ThoRusMor-22, ThoRusRem-24}. Two-colorability is also chosen together with the allocation of the qubits to the two tiers of Sec.~\ref{sec:03-A}.

Let $G=(V,E)$ be the graph associated with the distributed two-colorable resource state. Its vertex set is partitioned in two disjoint subsets $V_o = \{o_1, \dots ,o_{n_o}\}$ and $V_c = \{ c_1, \dots, c_\kappa \}$, where $|V_o| = n_o$ denotes the number of \textit{orchestration qubits} retained by the Tier2 node(s) and $|V_c| = \kappa$ represents the number of \textit{peer qubits} distributed to Tier1 nodes. Accordingly, the total number of qubits is $n = n_o + \kappa$. The number of Tier1 nodes is at most $\kappa$ (one qubit per node), but can be smaller if some Tier1 nodes hold multiple peer qubits. 

Following~\cite{MazCalCac-25}, we characterize the considered resource family through a small set of structural parameters that determine the structure and the admissible entanglement transformations. In particular, we distinguish \textit{peer degree}, the \textit{bridge rank} and \textit{bridge degree}.

\begin{definition}[\textbf{$\mathbf{r}$-rank bridge}]
	\label{def:r_rank_bridge}
    An $\mathbf{r}$-rank bridge $b_i$ is a peer qubit adjacent to $\mathbf{r} > 1$ orchestration qubits. The set of $\mathbf{r}$-rank bridges is formally defined as: 
	\begin{align}
		\label{eq:r_rank_bridge}
        \mathcal{B}_\mathbf{r} &= \big\{ b_i \in V_c : | N_{b_i} \cap V_o | = \mathbf{r} \big\}, \; \text{with}\, \mathbf{r} > 1.
	\end{align}
\end{definition}

\begin{definition}[\textbf{Bridge degree}]
	\label{def:bridge_degree}
    The bridge degree $\kappa_{b,\mathbf{r}}^{(o_i)} \leq \kappa$ of an orchestration qubit $o_i \in V_o$ with respect to rank $\mathbf{r}$ is the number of $\mathbf{r}$-rank bridges $b_i \in \mathcal{B}_\mathbf{r}$ belonging to its neighborhood $N_{o_i}$:
	\begin{align}
		\label{eq:bridge_degree}
		\kappa_{b,\mathbf{r}}^{(o_i)} = | \mathcal{B}_\mathbf{r} ^{(o_i)}| =      
        |\mathcal{B}_\mathbf{r} \cap N_{o_i} |, \; \; o_i \in V_o,
	\end{align}
    For brevity, when $\mathbf{r}=2$, we write $\mathcal{B} \equiv \mathcal{B}_2$ and $\kappa_b^{(o_i)} \equiv \kappa_{b,2}^{(o_i)}$.
\end{definition}

\begin{definition}[\textbf{Peer degree}]
	\label{def:peer_degree}
    The peer degree $\kappa_{c}^{(o_i)} \leq \kappa$, of an orchestration qubit $o_i \in V_o$ is the number of adjacent peer qubits:
	\begin{align}
		\label{eq:peer_degree}
        \kappa_{c}^{(o_i)} &= | N_{o_i} \cap V_c|, \; \; o_i \in V_o. 
	\end{align}
\end{definition}
We focus on \emph{regular} resource structures, where the bridge rank $\mathbf{r}$ is fixed across the topology (thus, the subscript will be omitted for notation simplification) and each orchestration qubit has either minimum bridge degree $\hat{\kappa}_b$ or maximum bridge degree $\bar{\kappa}_b$. Formally:
\begin{align}
	\label{eq:kappa}
	\hat{\kappa}_b = \min_{o_i \in V_o} \{ \kappa_b^{(o_i)} \}, \qquad   
	\bar{\kappa}_{b} = \max_{o_i \in V_o} \{ \kappa_b^{(o_i)} \}.
\end{align}

One of the simplest graph states that can be defined according to these parameters is the \textit{chain graph state} \cite{MazCalCac-25}, which is characterized by a 1D linear graph state structure with $\kappa_c=2$, $\hat{\kappa}_b = 1$, and $\mathbf{r} = 2$.
The fundamental resource family considered in this work is the \textit{Generalized Tree-like} (GTL) graph state, obtained by fixing $\mathbf{r}=2$. 

\begin{definition}[\textbf{Generalized Tree-like (GTL)}]
    \label{def:GTL}
    An $n$-qubit GTL graph state is a tuple
    \begin{equation}
        \label{eq:GTL}
        \mathrm{GTL} = (G,\, \kappa_c,\, \hat{\kappa}_b), \quad
        \kappa_c, \hat{\kappa}_b \in \mathbb{N}^+, \;\; \kappa_c \geq 2\hat{\kappa}_b,
    \end{equation}
    where $G = (V,E)$ is a connected two-colorable graph with vertex bipartition $V = V_o \cup V_c$, satisfying the following structural constraints:
    \begin{itemize}
        \item[\textbf{C1.}] \emph{(Peer degree regularity)} $|N_{o_i} \cap V_c| = \kappa_c$ for every $o_i \in V_o$, with $\kappa_c$ the peer degree in Def.~\ref{def:peer_degree}.
        \item[\textbf{C2.}] \emph{(Linear bridge ordering)} $V_o$ carries a linear order, every bridge $b \in \mathcal{B}$ (Def.~\ref{def:r_rank_bridge}) is adjacent to exactly one consecutive pair $(o_j, o_{j+1})$.
        \item[\textbf{C3.}] \emph{(Bridge regularity)} each consecutive pair $(o_j, o_{j+1})$ shares exactly $\hat{\kappa}_b$ bridges, with $\hat{\kappa}_b$ the minimum bridge degree in Def.~\ref{def:bridge_degree}.
    \end{itemize}
\end{definition}
From \textbf{C2}--\textbf{C3}, boundary orchestration qubits ($o_1$, $o_{n_o}$) have bridge degree $\hat{\kappa}_b$, and each interior vertex, when present, has bridge degree $2\hat{\kappa}_b$. For $n_o=2$ both vertices are boundary vertices and $\bar{\kappa}_b=\hat{\kappa}_b$. Hence, a GTL graph is uniquely determined, up to a relabeling of its vertices, by $n_o$, $\kappa_c$ and $\hat{\kappa}_b$, being formed by a linear backbone of orchestration qubits, where each consecutive pair shares $\hat{\kappa}_b$ bridges and each $o_i$ is further adjacent to $\kappa_c - \kappa_b^{(o_i)}$ peer qubits of degree one. Moreover, the number of orchestration qubits is $n_o = \frac{\kappa - \hat{\kappa}_b}{\kappa_c - \hat{\kappa}_b}$, with $n = n_o + \kappa$.

Within the configuration space induced by the GTL resource family, useful entanglement-connectivity patterns should target the following practically meaningful goals:
\begin{itemize}
    \item \textit{Entanglement link formation} -- enabling entanglement links between non-neighbor endpoints, by reducing the number of intermediate hops separating them in the resource graph;
    \item \textit{Parallel resource instantiation} -- maximizing the number of disjoint resources  (Bell pairs or GHZs or other communication resources) that can be concurrently instantiated from the shared resource;
    \item \textit{Hierarchy compliance} -- any realization mechanism must operate within the hierarchical responsibility model, with configurable operations confined to Tier2 nodes and requiring no inter-node coordination among Tier1 nodes.
\end{itemize}
The regime $\kappa_c > 2\hat{\kappa}_b$ yields additional non-bridge peer qubits that do not increase the number of concurrently instantiable entangled resources, offering no practical advantages for the above goals. Thus, for the remainder of this manuscript, we focus on the subclass $\kappa_c = 2\hat{\kappa}_b$. Under this restriction $\kappa_c$ is fully determined by $\hat{\kappa}_b$, and the GTL reduces to the two-parameter tuple
\begin{equation}
    \label{eq:GTL_specialized}
    \mathrm{GTL} = (G,\, \hat{\kappa}_b), \quad \hat{\kappa}_b \in \mathbb{N}^+,\;\; \hat{\kappa}_b \geq 2,
\end{equation}
for which the number of orchestration qubits is $n_o = (\kappa - \hat{\kappa}_b)/\hat{\kappa}_b$.

In the following, we show how to achieve all three aforementioned goals within the induced configuration space via a mechanism termed Entanglement Rolling.

\section{Entanglement Rolling}
\label{sec:04}

Here, we formalize Entanglement Rolling as a systematic procedure to navigate through the configuration space of the GTL resource family to engineer entanglement-graph configurations.

\subsection{Proximity and bridge neighborhoods}
\label{sec:04-A}

Sharing tailored resource states in a quantum network mitigates the limitations imposed by the physical network graph: even if endpoints are not connected by a physical link, they may become neighbors in the overlaid entanglement graph induced by the shared resource. In the GTL family, this entanglement graph is two-colorable, with bipartition $V=V_o\cup V_c$. Hence, peer qubits $c_i,c_j\in V_c$ are never adjacent, and any path between them necessarily alternates between peer and orchestration vertices. Accordingly, to quantify how ``far'' two peer qubits are, we measure their distance, following~\cite{MazCalCac-25}, in terms of the \textit{bridges} encountered along a shortest path. Cor.~\ref{cor:proximity_reduction} shows that this metric directly matches the number of Entanglement Rolling steps needed to establish an entangled link between the selected peer vertices.

\begin{definition}[\textbf{Peer proximity}]
	\label{def:peer_proximity}
    The peer proximity $\pi(c_i,c_j)$ between two peer qubits $c_i, c_j \in V_c \subset V$ with $i \neq j$, is defined as one plus the number of bridges on a shortest path $p_{c_i,c_j}$, between them in $G = (V, E)$:
	\begin{align}
		\label{eq:peer_proximity}
        \pi (c_i,c_j) &=  1+|\beta_{c_i,c_j}|,
	\end{align}
	with $\beta_{c_i,c_j} = \big\{ b \in p_{c_i, c_j} \setminus \{c_i, c_j\} : b \in \mathcal{B} \big\}$ denoting the set of bridges along the shortest path $p_{c_i,c_j}$, excluding its endpoints $c_i$ and $c_j$.
\end{definition}
To better capture the engineering role of bridges in our framework, we define the \textit{bridge neighborhoods}, which partition the peers adjacent to an orchestration vertex into two sets.
\begin{definition}[\textbf{Bridge neighborhoods}]
    \label{def:bridge_neighborhoods}
    Given an orchestration vertex $o_i \in V_o$, its neighborhood $N_{o_i}$ is partitioned into two disjoint sets: the left bridge neighborhood $\mathcal{L}_B^{(o_i)}$ and the right bridge neighborhood $\mathcal{R}_B^{(o_i)}$, defined as
    \begin{align}
        \label{eq:right_bridge_neighborhood}
        \mathcal{R}_{B}^{(o_i)} &{}= \mathcal{B}^{(o_i)} \cap \mathcal{B}^{(o_{i+1})}, \quad \text{if } o_{i+1} \in V_o,
    \end{align}
    \begin{align}
        \label{eq:left_bridge_neighborhood}
        \mathcal{L}_{B}^{(o_i)} &{}= \mathcal{B}^{(o_i)} \cap \mathcal{B}^{(o_{i-1})}, \quad \text{if } o_{i-1} \in V_o.
    \end{align}
    By C2 of Def.~\ref{def:GTL}, $\mathcal{L}_B^{(o_i)}$ and $\mathcal{R}_B^{(o_i)}$ are disjoint, and by C3 each has cardinality $\hat{\kappa}_b$. For the end orchestration qubits, where no bridge exists on one side, we denote by $\mathcal{R}_B^{(o_{n_o})} = N_{o_{n_o}} \setminus \mathcal{B}$ and $\mathcal{L}_B^{(o_1)} = N_{o_1} \setminus \mathcal{B}$ the sets of their peers that are not bridges, which also have cardinality $\hat{\kappa}_b$ since $\kappa_c = 2\hat{\kappa}_b$ and $n_o \geq 2$. Moreover, $\mathcal{R}_B^{(o_i)} = \mathcal{L}_B^{(o_{i+1})}$ for $i < n_o$, so that the $n_o+1$ bridge neighborhoods partition $V_c$.
\end{definition}

\begin{figure}
    \centering
    \includegraphics[width=0.8\linewidth]{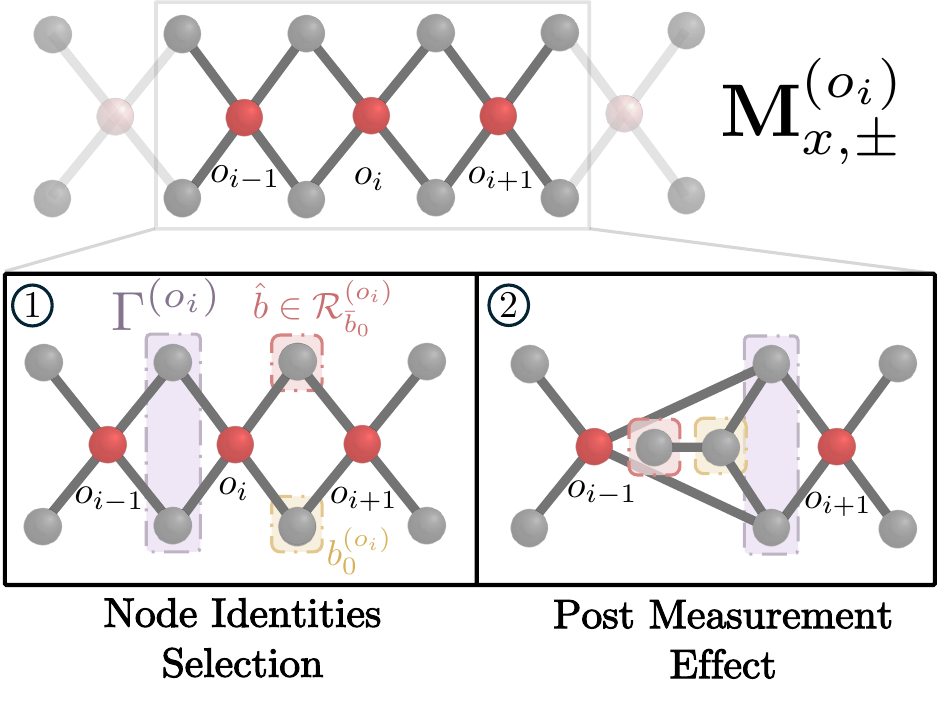}
    \caption{Illustration of the Entanglement Rolling effect induced by a Pauli-X measurement on orchestration qubit $o_i$ of a GTL resource state $(G,\hat{\kappa}_b=2)$ with arbitrary length.}
    \label{fig:03}
\end{figure}
\subsection{Entanglement Rolling Effect}
\label{sec:04-B}
Theorem~\ref{th:entanglement_rolling_effect} formalizes the \textit{Entanglement Rolling Effect} induced by a Pauli-X measurement on an orchestration qubit $o_i$ with a freely chosen support vertex $b_0^{(o_i)}$. This single-step effect is the elementary building block of the \textit{Entanglement Rolling procedure}: repeating this operation over a subset of the $n_o$ orchestration qubits, according to a configurable support vertex sequence $\mathcal{S}_{b_0} = \{b_0^{(o_1)}, \dots, b_0^{(o_{n_o})}\}$, 
yields the \emph{Entanglement Rolling procedure}, which progressively reduces the entanglement proximity among selected endpoints.

\begin{theorem}[Entanglement Rolling Effect]
    \label{th:entanglement_rolling_effect}
    Given a distributed $n$-qubit GTL graph state in a quantum network with $n_o$ orchestration qubits and $\kappa$ peer qubits, a Pauli-X measurement -- described by the manipulation operator $\mathbf{M}_{x,\pm}^{(o_i)}$ -- is applied on the orchestration qubit $o_i \in V_o$ with support vertex $b_0^{(o_i)}$$\,\in \mathcal{R}_B^{(o_i)}$. The measurement induces an Entanglement Rolling effect:
    \begin{itemize}
        \item[i.)] The designated vertex $b_0^{(o_i)}$ becomes the center of a star  connecting all neighbors of the measured vertex $o_i$.
        \item[ii.)] Peer qubits in $\Gamma^{(o_i)}= N_{o_i} \setminus \mathcal{R}_{B}^{(o_i)}$  become neighbors of the next orchestration qubit $o_{i+1}$, whenever $o_{i+1}$ exists.
    \end{itemize}
    \begin{proof}
        Please refer to Appendix~\ref{sec:Th1} for the proof.
    \end{proof}   
\end{theorem}

\begin{remark}
    \label{rem:left_support}
    Th.~\ref{th:entanglement_rolling_effect} is stated for $b_0^{(o_i)} \in \mathcal{R}_B^{(o_i)}$, but holds symmetrically for $b_0^{(o_i)} \in \mathcal{L}_B^{(o_i)}$ by reversing the ordering of orchestration qubits. Within the left-to-right ordering, if $b_0^{(o_i)} \in \Gamma^{(o_{i-1})}$ (with $\Gamma^{(o_0)} = \mathcal{L}_B^{(o_1)}$), the other vertices of $\Gamma^{(o_{i-1})}$ become adjacent to $b_0^{(o_i)}$ only, while $\mathcal{R}_B^{(o_i)}$ becomes the rolled set $\Gamma^{(o_i)}$ (Appendix~\ref{sec:Th1}).
\end{remark}

A pictorial representation of the results in Th.~\ref{th:entanglement_rolling_effect} is provided in Fig.~\ref{fig:03}, where $b_0$ and the set of \textit{rolled vertices} $\Gamma^{(o_i)}$ are highlighted before and after the measurement.
A direct consequence of Th.~\ref{th:entanglement_rolling_effect} is the reduction of entanglement proximity, not only for a target vertex $c_j$, but also for all vertices selected as support vertices. This is a key feature of the Entanglement Rolling effect, since it can be exploited to properly generate desired entanglement links by appropriately selecting the support vertices. This proximity reduction was first proved in~\cite{MazCalCac-25} for a fixed choice of the support vertices. We restate and prove it here to keep the paper self-contained, as it underpins the subsequent configuration-space and resource-instantiation analysis. Indeed, the proximity-reduction result follows from Th.~\ref{th:entanglement_rolling_effect}, by assuming a constant set of rolled vertices $\Gamma^{(o_i)}$ at each measurement step. 

\begin{corollary}[Entanglement Rolling: Proximity Reduction]
    \label{cor:proximity_reduction}
    Consider a distributed $n$-qubit GTL graph state with $\kappa$ peer qubits. Two peer qubits $c_i,c_j \in V_c$, $i \neq j$, can be made adjacent in the entanglement graph by performing $\pi(c_i,c_j)$ Pauli-X measurements on orchestration qubits belonging to the shortest path $p_{c_i,c_j}$ connecting $c_i$ and $c_j$ in $G$.

    \begin{proof}
        The proof follows by iterating $\pi(c_i,c_j)$ Pauli-X measurement steps described in Th.~\ref{th:entanglement_rolling_effect}, with $\pi(c_i,c_j)$ defined in Def.~\ref{def:peer_proximity}. At each step the rolled set becomes adjacent to the subsequent orchestration vertex, if any. By choosing the support-vertex sequence $\mathcal{S}_{b_0}$ such that the rolled set remains constant at every measurement step, i.e., $\Gamma = \Gamma^{(o_i)}, \forall o_i \in V_o$, the proximity $\pi(c_i,c_j)$ decreases by 1 at each measurement until the last measurement makes $c_i$ and $c_j$ adjacent in the entanglement graph. For an alternative derivation, please refer to~\cite{MazCalCac-25}.
    \end{proof}
\end{corollary}
Entanglement Rolling allows us to characterize and systematically achieve a subset of the full configuration space $\mathcal{G}$, defined as follows.

\begin{definition}[\textbf{Configurations Attainable by Entanglement Rolling}]
    \label{def:rolling}
    Given a GTL graph state $\ket{G}$ satisfying Eq.~\eqref{eq:GTL_specialized}, the set of configurations attainable by Entanglement Rolling when the orchestration qubits are measured, from $o_1$ to $o_{n_o}$, is
    \begin{equation*}
        \mathcal{G}_{\mathrm{ER}} = \bigl\{ \ket{G_{\mathcal{S}_{b_0}}} : b_0^{(o_i)} \in N_{o_i} \ \forall o_i \in V_o \bigr\} \subseteq \mathcal{G},
    \end{equation*}
    where $N_{o_i}$ is the neighborhood of $o_i$ when it is measured, and $\ket{G_{\mathcal{S}_{b_0}}}$ is the graph state reached by the Pauli-X measurements of the orchestration qubits with support sequence $\mathcal{S}_{b_0}$, in which each support vertex is the center of a star on the other peers of its bridge neighborhood and on the peers of the final rolled set, i.e., the bridge neighborhood without support vertices (Eqs.~\eqref{eq:C_2}--\eqref{eq:C_3}, Appendix~\ref{sec:Cor}).
\end{definition}

\begin{remark}
    Right supports at every step leave $\mathcal{L}_B^{(o_1)}$ as the final rolled set, whereas a support in $\mathcal{L}_B^{(o_1)}$ at step $j$, with right supports at all the other steps, leaves $\mathcal{R}_B^{(o_j)}$; in both cases, the center of every other group can be prescribed. The final graph identifies the rolled set as the only group without internal edges and each center as the unique vertex of its group adjacent to it. Thus the number of reachable labeled peer graphs is $|\mathcal{G}_{\mathrm{ER}}| = (n_o+1)\,\hat{\kappa}_b^{\,n_o}$.
    These configurations are related by local complementations of their graphs, i.e., equivalent~\cite{HeiDurEis-06}, and are all reachable with the same measurements. Unlike the Pauli-Y and Pauli-Z measurements, whose graph rules are unique, the Pauli-X measurement leaves the support vertex free~\cite{HeiDurEis-06}, and this choice is an additional layer of configurability, which makes multiple equivalent configurations reachable with the same type of operation.
\end{remark}

\section{Whatever Channel Resolution via Entanglement Rolling}
\label{sec:05}
This section completes the \textit{whatever-channel resolution} pipeline introduced in Sec.~\ref{sec:03}. In our framework, resolution is a structured LOCC process that selects a target entanglement-graph configuration from the admissible space induced by the resource and then realizes it. Entanglement Rolling implements the Tier2-side navigation of this space, while a second stage of local Tier1 operations isolates the desired end-to-end resources. 

\subsection{Two-Stage Resolution}
\label{sec:05-A}
\begin{figure}
    \centering
    \includegraphics[width=0.75\linewidth]{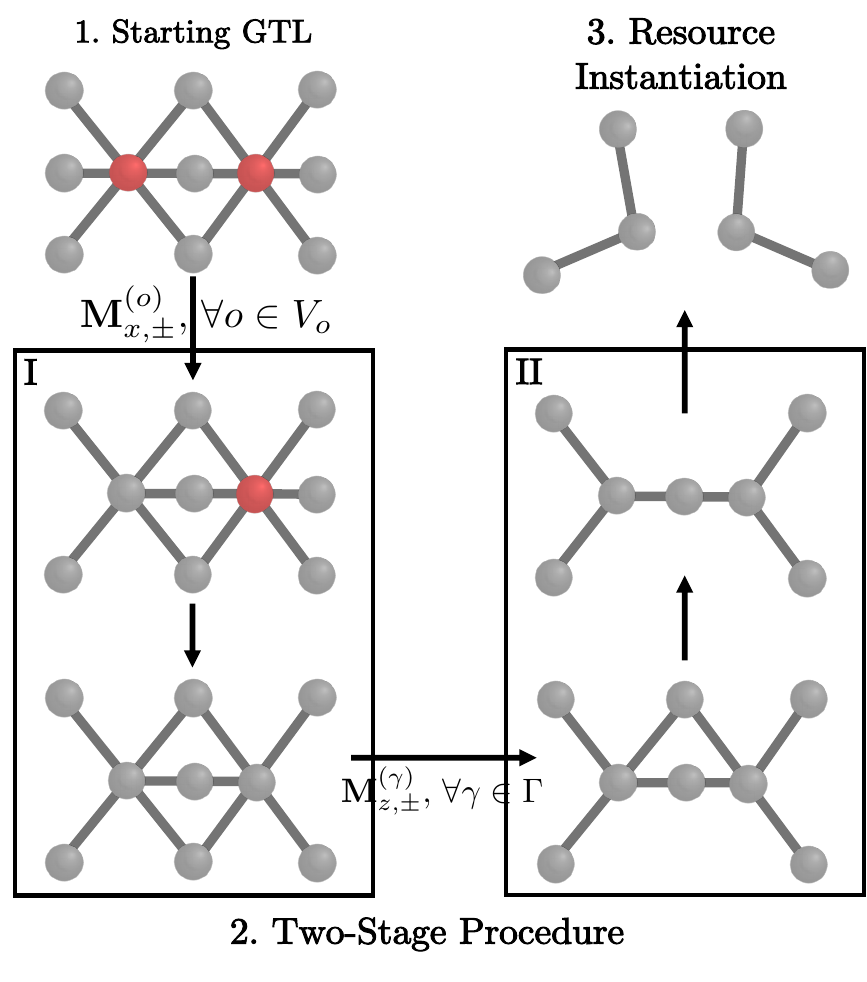}
    \caption{ 
    Example resource state extraction on a GTL resource state $(G,\,\hat{\kappa}_b = 3)$: Entanglement Rolling, i.e., a configurable sequence of Pauli-X measurements at Tier2, followed by local Pauli-Z measurements at the Tier1 end-node level, extracts two disjoint entangled resources (LU equivalent to 3-qubit GHZ states).
    }
    \label{fig:rolling_example}
\end{figure}

When a GTL graph state is distributed in the network, resolution proceeds in two stages as an LOCC process that instantiates a target entanglement-graph configuration from the admissible space, as depicted in Fig.~\ref{fig:rolling_example}. 

\textit{Stage~1 (Tier2 configuration selection).} Tier2 node(s) apply a configurable sequence of Pauli-X measurements on orchestration qubits in $V_o$, implementing Entanglement Rolling and steering the shared resource toward a target entanglement-graph configuration (Th.~\ref{th:entanglement_rolling_effect}, Cor.~\ref{cor:proximity_reduction}). Depending on which orchestration qubits are measured, the instantiated configuration may be \textit{pure-peer} (an entanglement graph supported only on $V_c$) or \textit{hybrid} (retaining a subset of unmeasured orchestration qubits and thus retaining Tier2 vertices in the instantiated entanglement graph).

Once a configuration has been instantiated, further LOCC manipulations can be applied \emph{on top} of it to realize specific entanglement-based functionalities. In particular, when the objective is to isolate end-to-end links between peer endpoints, \textit{Stage~2} applies a local \textit{isolation step} at Tier1 nodes: selected peer qubits are measured in the Pauli-Z basis to remove undesired vertices and obtain disjoint entanglement resources (e.g., Bell pairs or GHZ states) from the instantiated configuration. These operations are purely local and require no inter-node quantum coordination among Tier1 nodes, which rely on classical side information produced during Stage~1 and handled by the control plane~\cite{CacCalIll-26, CalCac-25}.

In the remainder of this section, we focus on pure-peer resolutions followed by Tier1-$Z$ stage, for isolating the desired end-to-end resources.

\begin{remark}[Architectural trade-off]
An alternative first stage, referred to as Y/Z resolution, performs at Tier2 $n_o$ Pauli measurements, in $Y$ at odd-indexed and $Z$ at even-indexed orchestration qubits, which split the peers into at most $\lceil n_o/2 \rceil$ disjoint entangled states, each among the neighbors of an orchestration qubit measured in the Pauli-Y basis. For chain resources ($\hat{\kappa}_b = 1$), these states are Bell pairs and all the measurements are concentrated at Tier2~\cite{MazCalCac-25}, whereas for $\hat{\kappa}_b \geq 2$ they are LU-equivalent to $2\hat{\kappa}_b$-qubit GHZ states, and Tier1 measurements are still required to isolate Bell pairs. This alternative comes at the cost of reduced reconfigurability, since the achievable patterns are constrained to peers at entanglement proximity $\pi = 1$ in the initial resource state. In contrast, our two-stage design trades a lightweight local Tier1 post-processing for increased configurability, enabling the systematic instantiation of a broader set of entanglement-graph configurations.
\end{remark}

In our approach, Tier1 nodes receive two types of classical messages.
(i) After Stage~1, Tier2 disseminates outcome-dependent correction instructions to the peer qubits in $N_{o_i}$, for each $o_i$; the designated support vertex $b_0^{(o_i)}$ receives a different correction than the remaining neighbors~\cite{HeiDurEis-06, MazZhaChu-25}. This asymmetry identifies the chosen support vertex $b_0^{(o_i)}$ and can be handled via standard Pauli-frame tracking, since the corrections of the two outcomes differ by a Pauli operator, while their non-Pauli part, acting on $b_0^{(o_i)}$ only, is fixed by the support sequence.
(ii) After the completion of Stage~1, Tier1 nodes receive \emph{isolation} directives specifying which additional local Pauli-Z measurements to perform to obtain the desired disjoint resources.

The selection of which resources to isolate (and when) depends on scheduling and policy decisions, including traffic patterns and control-plane objectives. These aspects are delegated to the control-plane/Tier2 interaction and a full treatment of such policies is left for future work. Here, we focus on the underlying resolution mechanism and assume that Tier1 nodes simply execute the received local directives without inter-node coordination. Overall, the sequence $\mathcal{S}_{b_0}=\{ b_0^{(o_1)}, \dots, b_0^{(o_{n_o})} \}$ captures the software-defined configuration of the whatever-channel resolution, while the non-support peers of $o_k$, $\mathcal{R}_{\bar b_0}^{(o_k)} = \mathcal{R}_B^{(o_k)} \setminus \{b_0^{(o_k)}\}$ or $\mathcal{L}_{\bar b_0}^{(o_k)} = \Gamma^{(o_{k-1})} \setminus \{b_0^{(o_k)}\}$ depending on the side of the support, form the set:
\begin{equation}
    \label{eq:non_b0_bridges}
    \mathcal{\bar S}_{b_0}^{(o_k)} =
    \begin{cases}
        \mathcal{R}_{\bar b_0}^{(o_k)} & \text{if } b_0^{(o_k)} \in \mathcal{R}_B^{(o_k)} \\
       \mathcal{L}_{\bar b_0}^{(o_k)} & \text{if } b_0^{(o_k)} \in \Gamma^{(o_{k-1})}
    \end{cases}.
\end{equation}

\begin{figure*}[t]
    \centering
    \begin{subfigure}[b]{0.65\textwidth}
        \centering
        \includegraphics[width=0.85\linewidth]{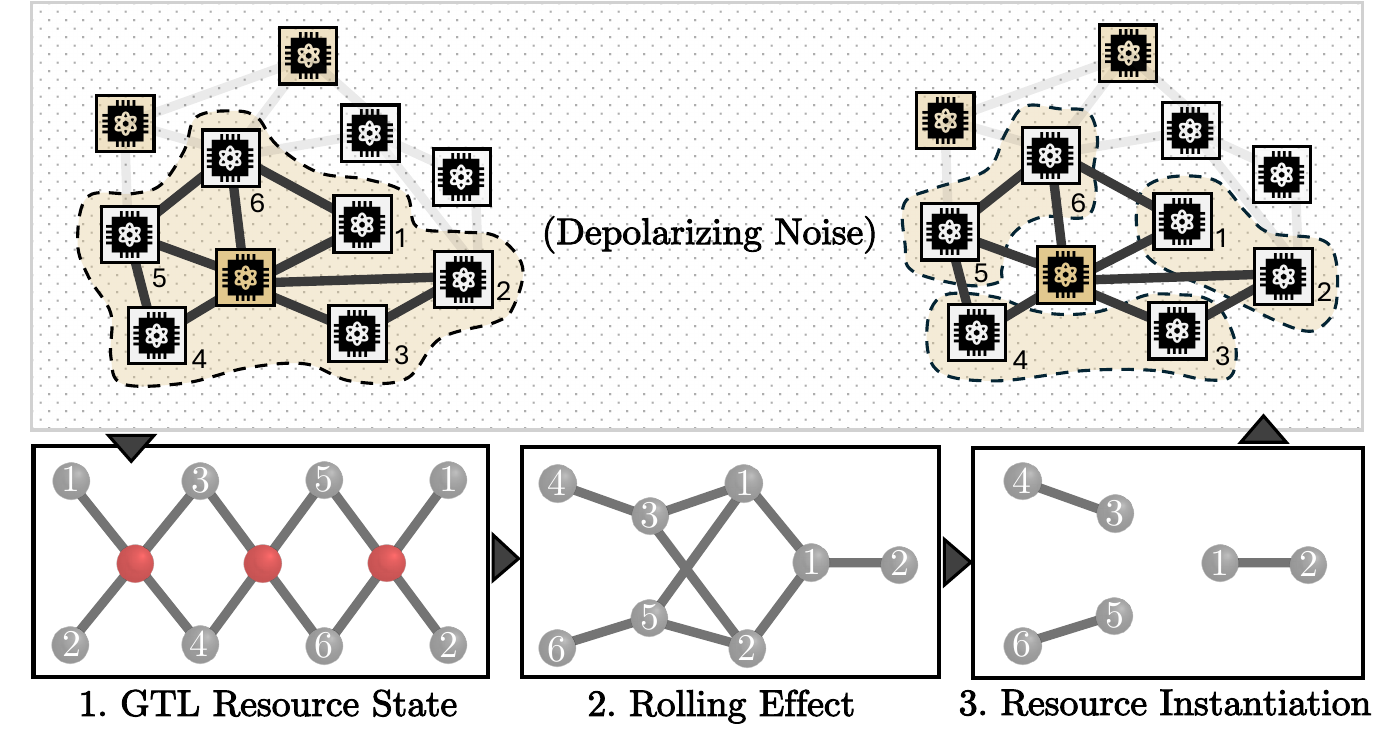}
        \caption{
        Noisy Entanglement Rolling and Bell pair instantiation.
        }
        \label{fig:noisy_rolling} 
    \end{subfigure}
    \begin{subfigure}[b]{0.3\textwidth}
        \centering
        \input{Tikz/BS_3_fidelity_non_time_dependent}
        \caption{
        Fidelity of the instantiated Bell pairs in presence of depolarizing noise.
        }
        \label{fig:fidelity_depolarizing}
    \end{subfigure}
    \caption{Example of (a) Entanglement Rolling with noisy GTL graph state $(G,\,\hat{\kappa}_b = 2)$. Node labels denote Tier1 nodes; repeated labels indicate that the same Tier1 node holds multiple peer qubits, as the general case discussed in Sec.~\ref{sec:03}. Pauli-X measurements at Tier2, followed by local Pauli-Z measurements at Tier1, isolate disjoint Bell pairs under depolarizing noise. (b) Fidelity of the extracted Bell pairs vs. the depolarizing parameter $p$ (one curve per extracted pair).}
\end{figure*}

\subsection{Representative Instance: Maximal parallel Bell Pairs}
\label{sec:05-B}
The proposed resource-driven framework supports multiple resolution objectives within the induced configuration space. As a key instance, we consider \textit{maximal parallel Bell-pair instantiation}, i.e., maximizing the number of disjoint Bell pairs that can be isolated concurrently from a distributed GTL state. Thanks to Entanglement Rolling, the maximum number of concurrently instantiable Bell pairs can be achieved by measuring (in Stage~1) the orchestration qubits via a suitable Pauli-X measurement sequence and then (in Stage~2) applying additional local Pauli-Z measurements at Tier1 nodes to isolate disjoint pairs, as formalized below.

\begin{corollary}[Entanglement Rolling: Maximal Parallel Bell-Pair Instantiation]
    \label{cor:maximal_ent_extraction}
    Consider a distributed $n$-qubit GTL graph state in a quantum network with $n_o$ orchestration qubits in the operating regime $\hat{\kappa}_b \geq 2$. It is possible to instantiate the maximum number $n_o$ of disjoint Bell pairs concurrently, by performing a total of $n_o$  Pauli-X measurements on the orchestration qubits, followed by $\hat{\kappa}_b + n_o(\hat{\kappa}_b - 2)$ local Pauli-Z measurements on peer qubits.
    \begin{proof}
        Please refer to Appendix~\ref{sec:Cor} for the proof.    
    \end{proof}
\end{corollary}
After Stage~1, the intermediate output is a collection of $n_o$ star graph states, centered in the support vertices and connected to each other through the $\hat{\kappa}_b$ peers of the rolled set $\Gamma^{(o_{n_o})}$ (Eqs.~\eqref{eq:C_2}--\eqref{eq:C_3}). The Pauli-Z measurements of these peers disconnect the stars (Eq.~\eqref{eq:C_4}), each LU-equivalent to a $\hat{\kappa}_b$-qubit GHZ state (see Fig.~\ref{fig:rolling_example}). For $\hat{\kappa}_b = 2$, the stars are then already LU-equivalent to Bell pairs and no further measurements are needed, i.e., two Tier1 Pauli-Z measurements in total. For $\hat{\kappa}_b \geq 3$, each star requires $\hat{\kappa}_b - 2$ additional Tier1 Pauli-Z measurements to isolate a Bell pair, i.e., $\hat{\kappa}_b + n_o(\hat{\kappa}_b - 2)$ in total, as in Cor.~\ref{cor:maximal_ent_extraction}. Cor.~\ref{cor:maximal_ent_extraction} thus quantifies parallel instantiation, with the center of each pair set by the support sequence $\mathcal{S}_{b_0}$ and its other endpoint chosen among the leaves in Stage~2, not the ability to serve an arbitrary matching among peers. Overall, the two-stage resolution measures each qubit outside the output Bell pairs exactly once, i.e., $n-2n_o$ measurements in total, and an exhaustive search over all single-qubit Pauli measurement patterns on $21$ GTL resources, with up to $31$ qubits, shows that no such pattern instantiates $n_o$ Bell pairs with fewer measurements. For the same number of Bell pairs, alternating Y/Z resolution requires $2n_o-1$ orchestration qubits at Tier2, compared with $n_o$ for Entanglement Rolling. On a chain, Y/Z resolution serves pairs at proximity one without Tier1 measurements, and their endpoints receive correction information one propagation delay after setup, instead of three for the two-stage resolution (Sec.~\ref{sec:06-B}). The companion code~\cite{repo} reproduces the matched comparison of qubit counts, measurements, correction latency and pair fidelities.

\section{Noise Analysis and Performance Evaluation}
\label{sec:06}

In this section, we analyze the proposed framework under physically motivated noise models. Leveraging the NSF introduced in Sec.~\ref{sec:02}, we derive closed-form noise maps that characterize how quantum noise propagates through the \textit{resource transformations} induced by Entanglement Rolling and by the subsequent local isolation steps. Importantly, these expressions depend on the local operations applied to the resource and on the noise models, rather than on how the orchestration qubits are partitioned across Tier2 nodes. Consequently, the derivation applies unchanged to deployments with one or multiple Tier2 nodes. Accordingly, for the performance evaluation, we instantiate the framework in a practically relevant tiered deployment in which a single Tier2 node distributes the GTL resource to multiple Tier1 nodes within a QLAN \cite{MazCalCac-25,MazZhaChu-25,PeaMazCal-26-journal}. This choice simplifies the distribution phase by construction -- which is not the focus of this paper -- while preserving the generality of the entanglement manipulation process supported by our framework.

We consider two main sources of noise acting on single qubits: (i) depolarizing Pauli noise $\mathcal{D}$, modeling the imperfections in resource generation and distribution, and (ii) a time-dependent dephasing noise $\mathcal{F}(t)$, modeling memory decoherence during the manipulation process, similarly to the approach in \cite{RuiWalDur-25}.

\subsection{Noisy Entanglement Manipulation}
\label{sec:06-A}
A depolarizing Pauli noise acting on qubit $a$ of a graph state $\varrho$ can be expressed in terms of Z-type operators only:
\begin{equation}
    \label{eq:depolarizing_channel}
    \mathcal{D}_a (\varrho) = p \varrho + \dfrac{1-p}{4}\sum_{\alpha,\beta \in \{0,1\}} (Z_a^{\alpha} \prod_{b \in N_a} Z_b^{\beta} ) \varrho (Z_a^{\alpha} \prod_{b \in N_a} Z_b^{\beta}).
\end{equation}
Its noise operators are $\mathbf{\Lambda}_{a,\alpha\beta} = Z_a^{\alpha}\prod_{b \in N_a} Z_b^{\beta}$, with $\alpha,\beta\in\{0,1\}$, and $p\in[0,1]$, where $p=1$ is the noiseless limit and $p=0$ the completely depolarizing channel.

The use of the NSF allows to update the noise operators after the application of manipulation operators. More specifically, if we consider the application of Entanglement Rolling or an arbitrary measurement strategy, the following manipulation operators are applied to the orchestration qubits: 
\begin{equation}
    \label{eq:manipulation_operators}
    \mathcal{O}_\xi = \{\mathbf{M}_{\xi,\pm}^{(o_1)}, \mathbf{M}_{\xi,\pm}^{(o_2)}, \dots \mathbf{M}_{\xi,\pm}^{(o_{n_o})}\}, \, \text{with } \xi \in \{x,y,z\},
\end{equation}
where the measurement operators are applied on the quantum state from left to right. If we consider that each qubit is independently affected by a depolarizing Pauli noise $\mathcal{D}_j$, the distributed noisy graph state is: $\mathcal{D}_1 \dots \mathcal{D}_n \ket{G}\bra{G}$, with $n=n_o+\kappa$. The NSF update rules reported in Tab.~\ref{tab:nsf_update_rules}, allow us to commute the noise through the measurement sequence, resulting in \textit{updated} local noise maps $\tilde{\mathcal{D}}_j$ acting on the post-measurement state. The next proposition provides closed-form expressions for the resulting updated depolarizing noise maps on peer and orchestration qubits.

\begin{proposition}
    \label{prop:noise_map_rolling}
    Consider a noisy GTL graph state $(G,\hat{\kappa}_b)$ with $\hat{\kappa}_b \geq 2$, affected by independent single-qubit depolarizing channels, and the ordered Entanglement-Rolling measurement sequence ${\mathcal{O}}_x=\{\mathbf{M}_{x,\pm}^{(o_1)},\dots,\mathbf{M}_{x,\pm}^{(o_{n_o})}\}$, with an arbitrary choice of the support vertices $b_0^{(o_i)}$. Using the NSF, one can update each local depolarizing channel separately. In particular, the updated depolarizing map associated with a noisy peer qubit $c_j\in N_{o_i}$ is given by:  
    \begin{align}
        \label{eq:peer_noise_map}
        \tilde{\mathcal{D}}_{c_j}(\varrho) &= p\varrho + 
        \tfrac{(1-p)}{4} \; \biggl[ \varrho + Z_{c_j} \, \varrho \, Z_{c_j} + \bigl( \prod_{k \in \tilde N} Z_k \bigr) 
        \; \varrho \; 
        \bigl( \prod_{k \in \tilde N} Z_k \bigr) \nonumber \\
        &+ \bigl( Z_{c_j} \prod_{k \in \tilde N} Z_k \bigr)
        \; \varrho  \;
        \bigl( Z_{c_j} \prod_{k \in \tilde N} Z_k \bigr) \biggr],
    \end{align}
    where $\varrho$ is the noiseless post-measurement graph state and the set $\tilde N$ depends on the type of peer qubit $c_j$:
    \begin{align}
        \label{eq:peer_neighborhood_noise}
        \tilde N = \begin{cases}
        \Gamma^{(o_{n_o})} \cup \mathcal{\bar S}_{b_0}^{(o_m)} & \begin{aligned} \text{if } &c_j = b_0^{(o_m)} \, , \\ &m\in\{1, \dots, n_o\} \end{aligned} \\
        \underset{\ell=1}{\bigcup^{n_o}} \
{b_0^{(o_\ell)}} & \text{if } c_j \in \Gamma^{(o_{n_o})} \\
        \{b_0^{(o_m)}\} & \begin{aligned} \text{if } &c_j \in \mathcal{\bar S}_{b_0}^{(o_m)}, \\ &m \in\{1, \dots, n_o\} \end{aligned}
    \end{cases}.
    \end{align}
Whereas the updated depolarizing map associated with a noisy orchestration qubit $o_i\in V_o$ is:   \begin{align}
        \label{eq:orchestrator_noise_map}
        \mathcal{\tilde D}_{o_i}(\varrho) =& p\varrho + \dfrac{(1-p)}{2} \cdot \\
        &\biggl[ \varrho + \bigl(\prod_{k \in \bar N } Z_k \bigr) \varrho \bigl(\prod_{k \in \bar N } Z_k \bigr)  \biggr], \nonumber
    \end{align}
    where the set $\bar N$ is given by:
    \begin{equation*}
        \bar N = \begin{cases}
            \bigcup_{\ell=i}^{n_o}\bigl\{b_0^{(o_\ell)}\bigr\} & \text{if } b_0^{(o_\ell)} \in \mathcal{R}_B^{(o_\ell)} \; \forall \ell \geq i \\
            \{b_0^{(o_j)}\} \cup \bigcup_{\ell=1}^{i-1}\bigl\{b_0^{(o_\ell)}\bigr\} & \text{otherwise}
        \end{cases},
    \end{equation*}
    with $j$ the first step $j \geq i$ such that $b_0^{(o_j)} \notin \mathcal{R}_B^{(o_j)}$.
\begin{proof}
    Please refer to Appendix~\ref{sec:noise_map_rolling} for the proof.
\end{proof}
\end{proposition}
The order of Pauli-X measurements affects the intermediate noise maps. We assumed the left-to-right ordering $\{o_1,\dots,o_{n_o}\}$; since the support vertices are never measured, permuting the fixed physical measurement pattern on the same noisy input, with the same final correction frame, preserves its output~\cite{RuiWalDur-25}. When all $n$ qubits are affected by noise, the final noisy state is obtained by composing the updated local maps over all the qubits, i.e., $\tilde{\mathcal{D}}_1 \tilde{\mathcal{D}}_2 \cdots\tilde{\mathcal{D}}_n \varrho$, with $\varrho$ the noiseless post-measurement graph state of Prop.~\ref{prop:noise_map_rolling}.

\begin{remark}
    By Tab.~\ref{tab:nsf_update_rules}, the Pauli-Z measurement of a peer $v$ deletes $Z_v$ from every noise operator. Since the support vertices are never measured, after Stage~2 the maps of the orchestration qubits are unchanged, each qubit of a Bell pair keeps Eq.~\eqref{eq:peer_noise_map} with $\tilde N$ reduced to the other qubit, and the map of each measured peer $c_j$ becomes
    \begin{equation*}
        \tilde{\mathcal{D}}_{c_j}(\varrho) = p\varrho + \dfrac{1-p}{2}\Bigl[\varrho + \bigl(\prod_{k \in \tilde N} Z_k\bigr)\,\varrho\,\bigl(\prod_{k \in \tilde N} Z_k\bigr)\Bigr],
    \end{equation*}
    with $\tilde N$ given by Eq.~\eqref{eq:peer_neighborhood_noise}, i.e., all the support vertices for the peers of $\Gamma^{(o_{n_o})}$, whose noise thus correlates the pairs.
\end{remark}

In words, the noise of a measured qubit is only partly removed with it: its errors that anticommute with the measured Pauli operator are transferred to the remaining qubits as phase flips. The instantiated resources are thus mixtures of the target state with states differing from it by phase flips, and their fidelity, given in closed form by Cor.~\ref{cor:bell_fidelity} for Bell pairs, quantifies the loss with respect to the noiseless case.
\subsection{Fidelity of Noisy Entanglement Resources}
\label{sec:06-B}
We now quantify the fidelity of the entanglement resources instantiated by whatever-channel resolution under noise. In the considered two-stage procedure, Entanglement Rolling selects a target entanglement-graph configuration, and local Pauli-Z measurements at Tier1 nodes isolate disjoint resources from the instantiated configuration, as illustrated in Figs.~\ref{fig:rolling_example} and~\ref{fig:noisy_rolling}. Building on the closed-form updated noise maps derived in Prop.~\ref{prop:noise_map_rolling}, we evaluate how depolarizing noise impacts GTL resource states and compute the fidelities\footnote{The experiments conducted in this section are openly available at~\cite{repo} and are based on the NSF library~\cite{nsf}.} for representative instantiated resources (Bell pairs and GHZ states). The fidelity of an instantiated resource is $F=\bra{\psi}\varrho\ket{\psi}$, where $\varrho$ is the reduced state of its qubits after both stages and $\ket{\psi}$ is the target graph state, i.e., the graph state of the edge between the two qubits of a Bell pair, or the star graph state, LU-equivalent to a GHZ state.

As depicted in Fig.~\ref{fig:noisy_rolling}, we consider the GTL graph state $(G,\,\hat{\kappa}_b = 2)$ with $n_o = 3$. By applying a two-stage resolution, it is possible to extract up to 3 disjoint, though correlated (Sec.~\ref{sec:06-A}), Bell pairs between end nodes. Moreover, in Fig.~\ref{fig:fidelity_depolarizing}, we plot the fidelity of each extracted Bell pair with respect to the depolarizing noise parameter $p$. At $p=0.98$, the three pairs have fidelities $0.942$, $0.933$ and $0.924$ in order of increasing proximity to the final rolled set. Their joint fidelity, the probability that all three are correct, is $0.869$, compared with $0.812$ for the product of their fidelities~\cite{repo}.

\begin{corollary}[Fidelity of the Instantiated Bell Pairs]
    \label{cor:bell_fidelity}
    Consider a noisy GTL graph state $(G,\hat{\kappa}_b)$ with $\hat{\kappa}_b\geq 2$, affected by independent single-qubit depolarizing channels, and the two-stage resolution of Cor.~\ref{cor:maximal_ent_extraction}, with an arbitrary choice of the support vertices $b_0^{(o_i)}$. The Bell pair instantiated between $b_0^{(o_m)}$ and the peer of $\mathcal{\bar S}_{b_0}^{(o_m)}$ not measured in Stage~2 has fidelity
    \begin{equation}
        \label{eq:bell_fidelity}
        F_m = \frac{1}{4}\Bigl(1 + p^2 + 2p^{\,2\hat{\kappa}_b+\pi(b_0^{(o_m)},\gamma)}\Bigr),
    \end{equation}
    with $\gamma$ any peer of $\Gamma^{(o_{n_o})}$ and $\pi$ the peer proximity of Def.~\ref{def:peer_proximity} in the initial resource state.
    \begin{proof}
        Please refer to Appendix~\ref{sec:bell_fidelity} for the proof.
    \end{proof}
\end{corollary}

The exponent counts the qubits whose noise reaches $b_0^{(o_m)}$, including the $\pi(b_0^{(o_m)},\gamma)$ orchestration qubits separating it from $\Gamma^{(o_{n_o})}$ in the initial resource state: the fewer they are, the higher the fidelity of the pair. Hence, the lowest fidelity is maximized by a final rolled set in the middle of the backbone, e.g., with a left support vertex at step $\lceil n_o/2 \rceil$ followed by right supports, and minimized by the all-right sequence adopted in our simulations.

To better describe the fidelity of the resulting states, we also take into account time-dependent dephasing noise:
\begin{equation}
    \label{eq:dephasing_channel}
    \mathcal{F}(t) = (1-q(t)) \varrho + q(t) Z \varrho Z, 
\end{equation}
where $q(t) = \frac{1}{2}(1-e^{-t/T})$ estimates the probability of a phase-flip error after a time $t$, and $T$ is the characteristic dephasing time of the quantum memory. Hence, the noise maps for each qubit $j$ of the graph state are given by the composition of the two noise channels, i.e., $\mathcal{F}_j(t_j) \mathcal{D}_j$, with $t_j$ the storage time of qubit $j$.
Therefore, the resulting noisy graph state is given by: $\mathcal{F}_1(t_1) \mathcal{D}_1 \dots \mathcal{F}_n(t_n) \mathcal{D}_n \ket{G}\bra{G}$. Since no measurement basis depends on earlier outcomes (Sec.~\ref{sec:05-A}) and $\mathcal{F}_j(t_j)$ commutes with the measurements of the other qubits, the dephasing of each qubit can be applied to the initial state and propagated by the rules of Tab.~\ref{tab:nsf_update_rules}.

\begin{figure*}
    \centering
    \begin{minipage}[t]{0.57\textwidth}
        \begin{subfigure}[b]{\linewidth}
            \centering
            \usetikzlibrary{patterns}
\pgfplotsset{every axis/.append style={font=\small, scaled ticks=false,
  axis line style={line width=1pt, color=black},
  tick style={line width=0.75pt, color=black}},%
  every colorbar/.append style={%
    axis line style={line width=0.55pt},
    tick style={line width=0.55pt}},%
  every axis title/.append style={yshift=-1.5ex, xshift=1ex}}
\begin{tikzpicture}

\definecolor{darkgray176}{RGB}{176,176,176}

\begin{groupplot}[group style={group size=2 by 1, horizontal sep=0.55cm},
width=4.75cm, height=3.5cm]
\nextgroupplot[
colormap={YlOrRd_r}{[1pt]
  rgb(0pt)=(0.502,0,0.149);
  rgb(250pt)=(0.890,0.102,0.110);
  rgb(500pt)=(0.992,0.553,0.235);
  rgb(750pt)=(0.996,0.851,0.463);
  rgb(1000pt)=(1,1,0.800)
},
point meta max=1,
point meta min=0.25,
tick align=outside,
tick pos=left,
ymode=log,
title={Best Bell pair ($\pi=1$)},
x grid style={darkgray176},
xlabel={Probability $p$},
xmin=0.7, xmax=1,
xtick style={color=black},
xtick={0.7,0.8,0.9,1.0},
xticklabels={
  \(\displaystyle {0.7}\),
  \(\displaystyle {0.8}\),
  \(\displaystyle {0.9}\),
  \(\displaystyle {1.0}\)
},
y grid style={darkgray176},
ylabel={Dephasing time (ms)},
ylabel style={yshift=-0.2cm},
ymin=1e-03, ymax=1e-01,
ytick style={color=black},
ytick={1e-03,1e-02,1e-01},
yticklabels={
  \(\displaystyle {1}\),
  \(\displaystyle {10}\),
  \(\displaystyle {100}\)
}
]
\addplot graphics [includegraphics cmd=\pgfimage,xmin=0.7, xmax=1, ymin=1e-03, ymax=1e-01] {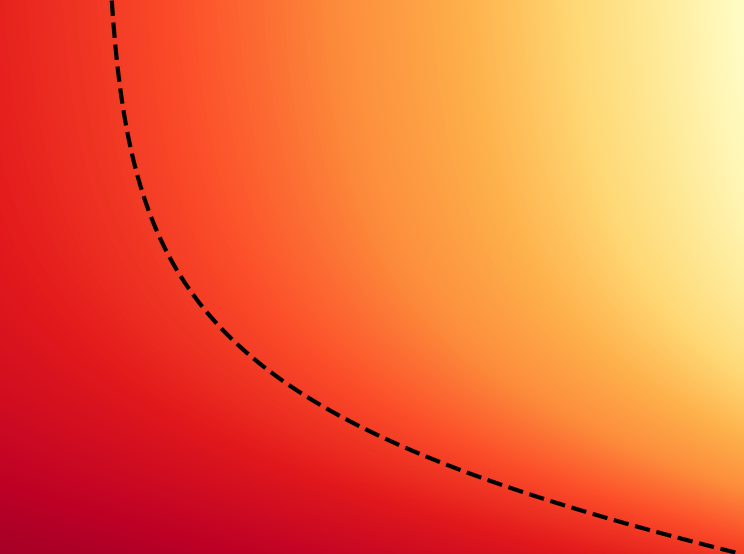};

\nextgroupplot[
colorbar,
colorbar style={width=0.5cm,xshift=-0.5cm,ytick={0.25,1},yticklabels={
  \(\displaystyle {0.25}\),
  \(\displaystyle {1}\)
},ylabel={Fidelity},ylabel style={yshift=0.65cm}},
colormap={YlOrRd_r}{[1pt]
  rgb(0pt)=(0.502,0,0.149);
  rgb(250pt)=(0.890,0.102,0.110);
  rgb(500pt)=(0.992,0.553,0.235);
  rgb(750pt)=(0.996,0.851,0.463);
  rgb(1000pt)=(1,1,0.800)
},
point meta max=1,
point meta min=0.25,
tick align=outside,
tick pos=left,
ymode=log,
title={Worst Bell pair ($\pi=4$)},
x grid style={darkgray176},
xlabel={Probability $p$},
xmin=0.7, xmax=1,
xtick style={color=black},
xtick={0.7,0.8,0.9,1.0},
xticklabels={
  \(\displaystyle {0.7}\),
  \(\displaystyle {0.8}\),
  \(\displaystyle {0.9}\),
  \(\displaystyle {1.0}\)
},
y grid style={darkgray176},
ytick={1e-03,1e-02,1e-01},
yticklabels={,,},
ymin=1e-03, ymax=1e-01
]
\addplot graphics [includegraphics cmd=\pgfimage,xmin=0.7, xmax=1, ymin=1e-03, ymax=1e-01] {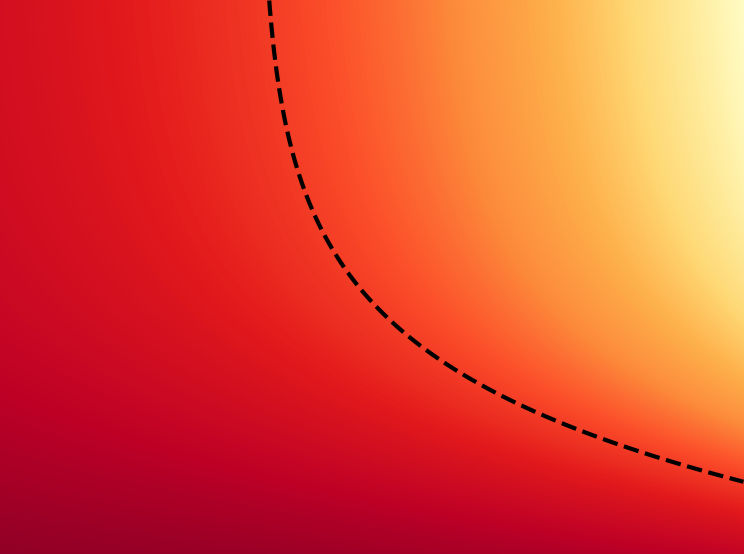};
\end{groupplot}

\end{tikzpicture}
            \caption{Extracted Bell-pair fidelity from $\mathrm{GTL}=(G,\hat{\kappa}_b{=}2)$ and $n_o{=}4$: best pair (left) and
            worst pair (right).}
            \label{fig:fidelity_BS_timedep}
        \end{subfigure}
        \\[1.5ex]
        \begin{subfigure}[b]{\linewidth}
            \centering
            \usetikzlibrary{patterns}
\pgfplotsset{
    every axis/.append style={
        font=\small, scaled ticks=false,
        axis line style={line width=1pt, color=black},
        tick style={line width=0.75pt, color=black}
    },
    every colorbar/.append style={%
        axis line style={line width=0.55pt},
        tick style={line width=0.55pt}
    },  
}

\begin{tikzpicture}
\definecolor{darkgray176}{RGB}{176,176,176}

\begin{groupplot}[group style={group size=2 by 1, horizontal sep=0.6cm},
width=4.75cm, height=3.5cm]
\nextgroupplot[
colormap={YlOrRd_r}{[1pt]
  rgb(0pt)=(0.502,0,0.149);
  rgb(250pt)=(0.890,0.102,0.110);
  rgb(500pt)=(0.992,0.553,0.235);
  rgb(750pt)=(0.996,0.851,0.463);
  rgb(1000pt)=(1,1,0.800)
},
point meta max=1,
point meta min=0.125,
tick align=outside,
tick pos=left,
ymode=log,
title={Best GHZ state ($\pi=1$)},
title style={yshift=-1.25ex},
x grid style={darkgray176},
xlabel={Probability $p$},
xmin=0.75, xmax=1,
xtick style={color=black},
xtick={0.75,0.85,0.95,1.0},
xticklabels={
  \(\displaystyle {0.75}\),
  \(\displaystyle {0.85}\),
  \(\displaystyle {0.95}\),
  \(\displaystyle {1.0}\)
},
y grid style={darkgray176},
ylabel={Dephasing time (ms)},
ylabel style={yshift=-0.2cm},
ymin=1e-03, ymax=1e-01,
ytick style={color=black},
ytick={1e-03,1e-02,1e-01},
yticklabels={
  \(\displaystyle {1}\),
  \(\displaystyle {10}\),
  \(\displaystyle {100}\)
}
]
\addplot graphics [includegraphics cmd=\pgfimage,xmin=0.75, xmax=1, ymin=1e-03, ymax=1e-01] {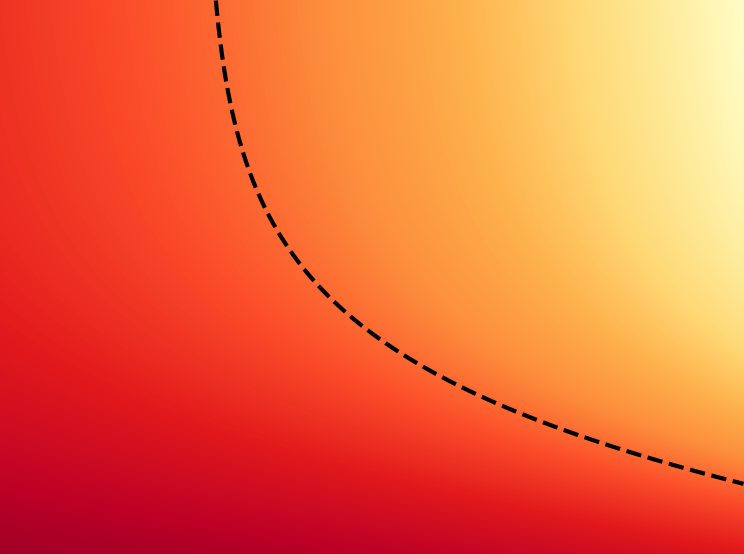};

\nextgroupplot[
colorbar,
colorbar style={width=0.5cm,xshift=-0.5cm,ytick={0.125,1},
yticklabels={
  \(\displaystyle {0.125}\),
  \(\displaystyle {1}\)
},ylabel={Fidelity},ylabel style={yshift=0.65cm}
},
colormap={YlOrRd_r}{[1pt]
  rgb(0pt)=(0.502,0,0.149);
  rgb(250pt)=(0.890,0.102,0.110);
  rgb(500pt)=(0.992,0.553,0.235);
  rgb(750pt)=(0.996,0.851,0.463);
  rgb(1000pt)=(1,1,0.800)
},
point meta max=1,
point meta min=0.125,
tick align=outside,
tick pos=left,
ymode=log,
title={Worst GHZ state ($\pi=4$)},
title style={yshift=-1.25ex},
x grid style={darkgray176},
xlabel={Probability $p$},
xmin=0.75, xmax=1,
xtick style={color=black},
xtick={0.75,0.85,0.95,1.0},
xticklabels={
  \(\displaystyle {0.75}\),
  \(\displaystyle {0.85}\),
  \(\displaystyle {0.95}\),
  \(\displaystyle {1.0}\)
},
y grid style={darkgray176},
ytick={1e-03,1e-02,1e-01},
yticklabels={,,},
ymin=1e-03, ymax=1e-01
]
\addplot graphics [includegraphics cmd=\pgfimage,xmin=0.75, xmax=1, ymin=1e-03, ymax=1e-01] {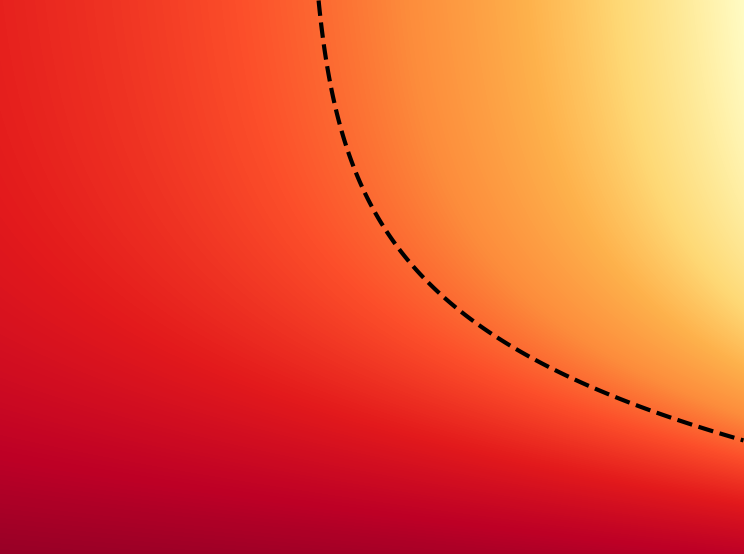};
\end{groupplot}

\end{tikzpicture}
            \caption{Extracted GHZ-state fidelity from $\mathrm{GTL}=(G,\hat{\kappa}_b{=}3)$ and $n_o{=}4$: best triple (left) and
            worst triple (right).}
            \label{fig:fidelity_GHZ_timedep}
        \end{subfigure}
    \end{minipage}%
    \hfill
    \begin{minipage}[t]{0.42\textwidth}
        \begin{subfigure}[b]{\linewidth}
            \centering
            \pgfplotsset{
    every axis/.append style={font=\footnotesize, scaled ticks=false,
    axis line style={line width=0.8pt, color=black},
    tick style={line width=0.75pt, color=black}},
}

\begin{tikzpicture}
\begin{axis}[
  width=6.05cm, height=3.55cm,
  font=\small,
  scaled ticks=false,
  ymode=log,
  unbounded coords=jump,
  tick align=outside,
  tick pos=left,
  xlabel={Probability $p$},
  ylabel={Dephasing time (ms)},
  ylabel style={yshift=-0.2cm},
  title={Fidelity threshold (worst Bell pair)},
  title style={yshift=-1.25ex},
  xmin=0.75, xmax=1.0,
  xtick={0.80,0.90,1.0},
  xticklabels={
    \(\displaystyle{0.8}\),
    \(\displaystyle{0.9}\),
    \(\displaystyle{1.0}\)
  },
  ymin=1e-3, ymax=1e-1,
  ytick={1e-3,1e-2,1e-1},
  yticklabels={$1$, $10$, $100$},
  legend pos=north east,
  legend style={font=\tiny, at ={(1,1)}},
]

\addplot[line width=1.5pt, color=blue]
  table[x=p, y=T, col sep=space] {Figures/fig6_threshold_bell_no2.dat};
\addlegendentry{$n_o=2$}

\addplot[line width=1.5pt, color=orange, dashed]
  table[x=p, y=T, col sep=space] {Figures/fig6_threshold_bell_no3.dat};
\addlegendentry{$n_o=3$}

\addplot[line width=1.5pt, color=green!60!black, dashdotted]
  table[x=p, y=T, col sep=space] {Figures/fig6_threshold_bell_no4.dat};
\addlegendentry{$n_o=4$}

\addplot[line width=1.5pt, color=red!70!black, dotted]
  table[x=p, y=T, col sep=space] {Figures/fig6_threshold_bell_no5.dat};
\addlegendentry{$n_o=5$}

\end{axis}
\end{tikzpicture}
            \caption{Fidelity $F{=}0.5$ threshold for Bell-pair extraction from $\mathrm{GTL}=(G,\hat{\kappa}_b{=}2)$ with $n_o \in \{2,3,4,5\}$.
            }
            \label{fig:fidelity_threshold_BS}
        \end{subfigure}
        \\[1.5ex]
        \begin{subfigure}[b]{\linewidth}
            \centering
            \pgfplotsset{
    every axis/.append style={font=\footnotesize, scaled ticks=false,
    axis line style={line width=0.8pt, color=black},
    tick style={line width=0.75pt, color=black}},
}

\begin{tikzpicture}
\begin{axis}[
  width=6.05cm, height=3.55cm,
  font=\small,
  scaled ticks=false,
  ymode=log,
  unbounded coords=jump,
  tick align=outside,
  tick pos=left,
  xlabel={Probability $p$},
  ylabel={Dephasing time (ms)},
  ylabel style={yshift=-0.2cm},
  title={Fidelity threshold (worst GHZ)},
  title style={yshift=-1.25ex},
  xmin=0.83, xmax=1.0,
  xtick={0.85,0.9,1.0},
  xticklabels={
    \(\displaystyle{0.85}\),
    \(\displaystyle{0.9}\),
    \(\displaystyle{1.0}\)
  },
  ymin=1e-3, ymax=1e-1,
  ytick={1e-3,1e-2,1e-1},
  yticklabels={$1$, $10$, $100$},
]
\addplot[line width=1.5pt, color=blue]
  table[x=p, y=T, col sep=space] {Figures/fig6_threshold_ghz_no2.dat};

\addplot[line width=1.5pt, color=orange, dashed]
  table[x=p, y=T, col sep=space] {Figures/fig6_threshold_ghz_no3.dat};

\addplot[line width=1.5pt, color=green!60!black, dashdotted]
  table[x=p, y=T, col sep=space] {Figures/fig6_threshold_ghz_no4.dat};

\addplot[line width=1.5pt, color=red!70!black, dotted]
  table[x=p, y=T, col sep=space] {Figures/fig6_threshold_ghz_no5.dat};

\end{axis}
\end{tikzpicture}
            \caption{Fidelity $F{=}0.5$ threshold for GHZ-state extraction from $\mathrm{GTL} = (G,\hat{\kappa}_b{=}3)$ with $n_o \in \{2,3,4,5\}$.
            }
            \label{fig:fidelity_threshold_GHZ}
        \end{subfigure}
    \end{minipage}
    \caption{Simulation of noisy entanglement extraction under depolarizing and time-dependent dephasing noise. Figs.~\ref{fig:fidelity_BS_timedep} and~\ref{fig:fidelity_GHZ_timedep} represent fidelity heatmaps for $n_o{=}4$ GTL resource states. The dashed black contour marks the $F{=}0.5$ threshold. Figs.~\ref{fig:fidelity_threshold_BS} and~\ref{fig:fidelity_threshold_GHZ} represent $F{=}0.5$ noise threshold curves for different GTL resource states with $n_o \in \{2,3,4,5\}$.}
    \label{fig:fidelity_results}
\end{figure*}
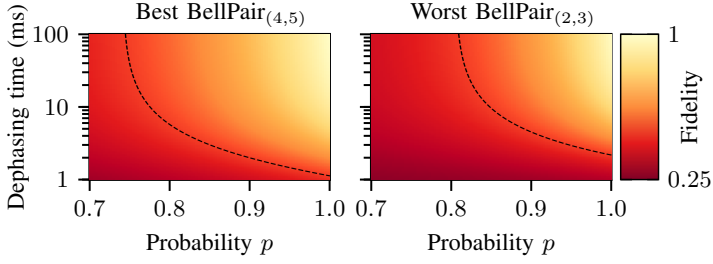
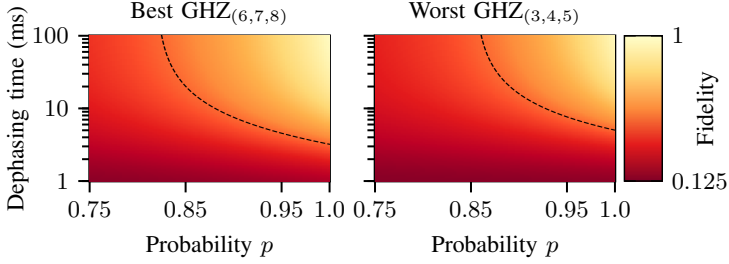
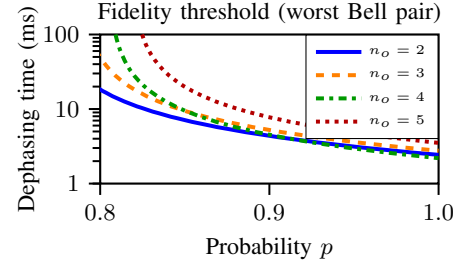
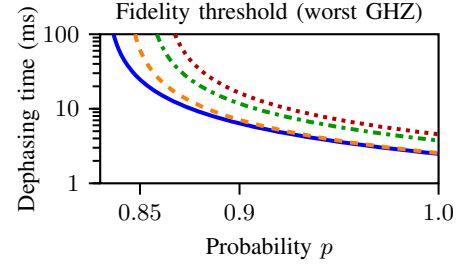

Figure~\ref{fig:fidelity_BS_timedep} gives the Bell-pair fidelities for $n_o=4$ and $\hat{\kappa}_b=2$ as functions of the depolarizing parameter $p$ and memory dephasing time $T$.
Tier2 dispatches the peers at time zero. At $40$\,km and $5\,\mu$s/km~\cite{PeaMazCal-26-journal}, the one-way delay is $\tau=0.2$\,ms. The peers arrive at $\tau$ and their acknowledgments reach Tier2 at $2\tau$. Tier2 selects the configuration, measures its qubits and sends the Stage~2 directives. Tier1 measures on receipt and returns the outcomes via Tier2 to the endpoints. Each stage uses parallel measurements and a software correction frame. With $1\,\mu$s per local round, orchestration qubits are stored for $0.401$\,ms from dispatch. Measured and retained peers are stored for $0.402$ and $0.803$\,ms from arrival, respectively. The endpoints receive correction information at $1.003$\,ms from dispatch.

Similarly, Fig.~\ref{fig:fidelity_GHZ_timedep} considers a GTL graph state $(G,\,\hat{\kappa}_b = 3)$ with $n_o = 4$, which can instantiate up to $4$ three-qubit GHZ states. We observe higher fidelity for the resources instantiated at the first steps of the measurement sequence than for the last ones: with the support vertices in the right bridge neighborhoods, as in Figs.~\ref{fig:fidelity_depolarizing} and~\ref{fig:fidelity_results}, $\Gamma^{(o_{n_o})}=\mathcal{L}_B^{(o_1)}$, and the support vertex of step $m$ is at proximity $m$ from it (Cor.~\ref{cor:bell_fidelity}). In both configurations, the fidelity remains above $F{=}0.5$ across a broad range of noise parameters. A fidelity strictly above this threshold suffices for distillability of both Bell pairs~\cite{BenBraPop-96} and GHZ states~\cite{DurCir-00}. As $n_o$ increases, more orchestration qubits contribute noise to the worst output. The above-threshold region narrows but persists for $n_o=5$.

\section[Conclusion and Outlook]{Conclusion and Outlook}
\label{sec:07}
In this work, we introduced a resource-driven framework for the programmable reconfiguration of multipartite entanglement, in line with the entanglement-defined networking vision underpinning emerging Quantum Internet architectures and protocol suites~\cite{CacCal-26, CalCac-25}. Rather than viewing entanglement manipulation solely as a tool to satisfy predefined connectivity requests, we modeled a shared multipartite entangled state as a \textit{whatever channel}, that can be systematically resolved via LOCC into different entanglement-connectivity configurations, from an admissible configuration space. Focusing on the GTL family, we formalized Entanglement Rolling as a measurement-based mechanism that systematically navigates this space. We proved that for GTL graph states with $\hat{\kappa}_b \geq 2$, Entanglement Rolling, followed by a local isolation step at Tier1, attains the maximum number of concurrently instantiable Bell pairs. Using the Noisy Stabilizer Formalism, we derived closed-form updated noise maps for Entanglement Rolling and the subsequent isolation steps, as well as the fidelity of the instantiated Bell pairs, and evaluated performance under depolarizing and time-dependent dephasing noise. Across a broad range of noise parameters, the resulting Bell-pair and GHZ-state fidelities remain above the threshold $F=0.5$, establishing operating regimes in which the extracted resources remain distillable under the considered noise models.

Besides the end-user applications recalled in Sec.~\ref{sec:01}, the resolution of the whatever channel may also serve the operation of the network itself, e.g., through entanglement-based switching fabrics for quantum forwarding~\cite{IllDeRCalCac-26}, or in entanglement-assisted interactions between control-plane functions and data-plane nodes. The classical signaling of the two resolution stages is governed by the control plane, whose architecture, protocols and optimizations for large-scale quantum networks are already being investigated~\cite{CacCal-26, CalCac-25, CacCalIll-26}. On the physical side, the generation at scale and the distribution of the resource state, abstracted in this work, remain the main challenges. Yet, photonic graph states have already been generated deterministically and fused~\cite{ThoRusRem-24}, also deterministically~\cite{LiYanBao-26}, reaching $13$ qubits, beyond the eight qubits of the smallest GTL resource that instantiates two Bell pairs, while the distribution can rely on the network-level mechanisms recalled in Sec.~\ref{sec:01-A}.

The framework is formulated in terms of graph states and single-qubit operations, which keeps the analysis platform-agnostic and analytically tractable. Once a GTL state is distributed, its resolution requires no entangling operation, and the most stringent requirement is the storage of the qubits of the instantiated resources until the outcomes that fix their corrections are received, since the other peers can be measured as soon as Tier2 establishes the configuration. This calls for quantum memories with coherence times compatible with the protocol duration of Sec.~\ref{sec:06-B} and, for photonic resources, for light--matter interfaces. Small-scale demonstrations of the resolution are therefore a plausible experimental direction, while its assessment under realistic network conditions, including loss, decoherence and coordination overheads, calls for simulation tools~\cite{MazZhaChu-25, PeaMazCal-26-journal} and testbeds.

Finally, our analysis navigates the configuration space through Entanglement Rolling on the GTL family, and evaluates it under independent depolarizing and dephasing noise. In this setting, the threshold $F = 0.5$ is a reference benchmark: above it, the instantiated Bell pairs and GHZ states are distillable~\cite{BenBraPop-96, DurCir-00}, although the yield of their purification depends on the protocol and on the full noisy state. The natural continuation of this work is the extension of the analysis to other resource families, to further imperfections, such as generation loss, gate and measurement errors and correlated noise, and to the scheduling and control-plane policies coordinating multiple Tier2 nodes. We hope that it will foster the design of multipartite resources as configurable whatever channels for future quantum networks.

\begin{appendices}

\section{Proof of Theorem 1} 
\label{sec:Th1}
We assume that a Pauli-X measurement is performed on orchestration qubit $o_i \in V_o$ and with corresponding $i$-th support vertex $b_0^{(o_i)}$ selected in the right bridge neighborhood for every measured qubit: $b_0^{(o_i)} = \bar r \in \mathcal{R}_{B}^{(o_i)}$.
As the choice is arbitrary, the proof can be carried similarly by choosing $b_0^{(o_i)}  = \bar \ell \in \mathcal{L}_{B}^{(o_i)}$.

When performing a Pauli-X measurement the effects on the resulting graph state -- up to local corrections -- are given by the graph $\tau_{b_0}\big(\tau_{o_i}(\tau_{b_0}(G)) - o_i\big)$. Specifically, after the local complementation $\tau_{b_0}(G)$, the resulting associated graph $G' = (V',E')$ can be expressed as follows:
        
\begin{equation}
\label{eq:T_3}
    G' = 
    \begin{cases}
        (V, E \cup \{\{o_i, o_{i+1}\}\}) & \text{if} \; \; i < n_o \\
        (V, E) & \text{if} \; \; i = n_o    
    \end{cases}.
\end{equation}
Moreover, starting from the graph $G'$ another local complementation $\tau_{o_i}(G')$ is performed leading to the graph $G''$ whose edge set $E''$ is given by: 
        
\begin{equation}
\label{eq:T_4}
    E''= (E' \cup {N_{o_i}'}^{2}) \setminus E_{N'_{o_i}} \; ,
\end{equation}
where, according to Eq.~\eqref{eq:T_3} the neighborhood $N'_{o_i}$ includes the subsequent orchestration qubit $o_{i+1}$, if $i < n_o$. In that case, local complementation at $o_i$ toggles the edge between each of its peer neighbors and $o_{i+1}$. Here, ${N_{o_i}'}^{2}$ denotes the set of unordered pairs of neighbors,
\begin{equation}
\label{eq:T_5}
    {N_{o_i}'}^{2} = \binom{N'_{o_i}}{2}.
\end{equation}
Accordingly, $E_{N'_{o_i}}$ includes the pre-existing edges between $o_{i+1}$ and the right bridges $\mathcal{R}_{B}^{(o_i)}$, i.e., every link of the kind $\{o_{i+1},r\}, \, \forall r \in \mathcal{R}_{B}^{(o_i)}$. 

As a consequence, these links are removed from $E''$ leading to the disconnections of the bridges $\mathcal{R}_{B}^{(o_i)}$ from orchestration qubit $o_{i+1}$, if any. Formally:
\begin{equation}
\label{eq:T_6}
    E_{N'_{o_i}} = 
    \begin{cases}
        \{ \{ o_{i+1}, r \}: r \in \mathcal{R}_{B}^{(o_i)} \} & \text{if} \; \; i < n_o \\
        \emptyset & \text{if} \; \; i = n_o
    \end{cases}.
\end{equation}

Furthermore, when $o_i$ is removed the resulting graph $G'''$ can be expressed as follows: 
\begin{equation}
\label{eq:T_9}
    G'''=G''-o_i.
\end{equation}
    
The last local complementation is, once again, applied on $b_0^{(o_i)}$. This leads to the graph $G^{\RN{4}}$, whose edge set is given by:
\begin{equation}
    \label{eq:T_10}
    E^{\RN{4}} = (E''' \cup {N_{b_0}'''}^{2}) \setminus E_{N'''_{b_0}}.
\end{equation}
    
After deletion, the neighbors of $b_0^{(o_i)}$ are the other peers in $N_{o_i}$. They form a clique in $G'''$. The last local complementation removes the edges of this clique and preserves the edges incident on the support, giving
\begin{equation}
    \label{eq:T_12}
    N^{\RN{4}}_{b_0^{(o_i)}} = N_{o_i} \setminus \{b_0^{(o_i)}\}.
\end{equation}
Hence, since no more edges are present between the vertices of the set $N_{o_i} \setminus \{b_0^{(o_i)}\}$, the current support vertex $b_0^{(o_i)}$ is also a star vertex for each of its neighbors, satisfying point i.) of the theorem.
    
According to Eq.~\eqref{eq:T_4}, every non $b_0^{(o_i)}$ right bridge (i.e., $\hat b \in \mathcal{R}_{\bar b_0}^{(o_i)}$) is no longer connected to $o_{i+1}$, thus losing its bridge role: $N^{\RN{4}}_{\hat b} = \{b_0^{(o_i)}\}, \; \; \forall \hat b \in \mathcal{R}_{\bar b_0}^{(o_i)}$.
Conversely, we denote the remaining qubits of the pre-measurement neighborhood $N_{o_i}$ as $\gamma \in \Gamma$, i.e. $\gamma \notin \mathcal{R}_{B}^{(o_i)}, \, \forall \gamma \in \Gamma$. 
Moreover, according to Eqs.~\eqref{eq:T_4}-\eqref{eq:T_5} we have that $\Gamma = N_{o_i} \setminus \mathcal{R}_{B}^{(o_i)}$ with neighborhood:
\begin{equation}
    \label{eq:T_14}
    N^{\RN{4}}_{\gamma} = 
    \begin{cases}
        (N_\gamma \setminus \{o_i\}) \cup \{b_0^{(o_i)}, o_{i+1}\} & \text{if} \; \; i < n_o \\
        (N_\gamma \setminus \{o_i\}) \cup \{b_0^{(o_i)}\} & \text{if} \; \; i = n_o
        \end{cases}.
    \end{equation}
    This means that each qubit $\gamma \in \Gamma$ has a direct link to $b_0^{(o_i)}$ and $o_{i+1}$, if any.
    In other words, the set of vertices $\Gamma$ replaces the set $\mathcal{R}_{B}^{(o_i)}$ in the resulting graph and becomes left bridge set $\mathcal{L}_{B}^{(o_{i+1})}$ for the subsequent orchestration qubit $o_{i+1}$, if any.
    Hence, point ii.) follows.

    The proof above uses only two properties of the graph before the measurement. First, the support vertex $b_0^{(o_i)}$ is adjacent only to $o_i$ and $o_{i+1}$, or only to $o_i$ if $i = n_o$. Second, no two vertices of $N_{o_i}$ are adjacent and, among them, only the vertices of $\mathcal{R}_B^{(o_i)}$ are adjacent to $o_{i+1}$. Along a support sequence, the graph before the measurement of $o_i$ is the GTL graph in which every measured $o_\ell$, $\ell<i$, is replaced by its support vertex $b_0^{(o_\ell)}$, the center of a star on $\mathcal{\bar S}_{b_0}^{(o_\ell)} \cup \Gamma^{(o_{i-1})}$, and in which the rolled set $\Gamma^{(o_{i-1})}$ takes the place of $\mathcal{L}_B^{(o_i)}$, with $\Gamma^{(o_0)} = \mathcal{L}_B^{(o_1)}$. For $b_0^{(o_i)} \in \mathcal{R}_B^{(o_i)}$, this gives both properties, which also hold along a run of right supports starting from another orchestration qubit, as in Cor.~\ref{cor:proximity_reduction}, and Eqs.~\eqref{eq:T_6}--\eqref{eq:T_14} give the same structure before the measurement of $o_{i+1}$, with the same rolled set. For $b_0^{(o_i)} \in \Gamma^{(o_{i-1})}$, the support vertex is also adjacent to the previous support vertices and, along a run starting from $o_k$, $k>1$, to $o_{k-1}$, which the three local complementations disconnect from $\Gamma^{(o_{i-1})}$ and connect to $\mathcal{R}_B^{(o_i)}$. Hence, $b_0^{(o_i)}$ is still the center of a star on $N_{o_i} \setminus \{b_0^{(o_i)}\}$, and $\mathcal{R}_B^{(o_i)}$ becomes the rolled set, as stated in the Remark in Sec.~\ref{sec:04-B}, giving again the same structure. Thus, Th.~\ref{th:entanglement_rolling_effect} and the Remark apply at every step of every support sequence.

\section{Proof of Corollary \ref{cor:maximal_ent_extraction}}
\label{sec:Cor}
By applying Th.~\ref{th:entanglement_rolling_effect} at each of the $n_o$ orchestration qubits in sequence, each $b_0^{(o_i)}$ vertex acquires the neighborhood given by Eq.~\eqref{eq:T_12}, with no edges among its neighbors. Specifically, every vertex of $\mathcal{\bar S}_{b_0}^{(o_i)}$ is no longer connected to any orchestration qubit, according to Th.~\ref{th:entanglement_rolling_effect} and the Remark after it. The total number of these vertices is:
\begin{equation}
    \label{eq:C_1}
    |\mathcal{\bar S}_{b_0}^{(o_i)}| = \hat{\kappa}_b - 1, \; \;\forall o_i \in V_o,
\end{equation}
independently of the choice of the support vertex $b_0^{(o_i)}$ at each measurement step $i$, since $|\mathcal{R}_B^{(o_i)}| = |\Gamma^{(o_{i-1})}| = \hat{\kappa}_b$ by Def.~\ref{def:bridge_neighborhoods}.

By the argument at the end of Appendix~\ref{sec:Th1}, the rolled set $\Gamma^{(o_{n_o})}$, with $|\Gamma^{(o_{n_o})}| = \kappa_c - \hat{\kappa}_b$, is connected to every designated $b_0^{(o_i)} \in \mathcal{S}_{b_0}$:
\begin{equation}
    \label{eq:C_2}
    N_{\Gamma^{(o_{n_o})}} = \{b_0^{(o_1)}, b_0^{(o_2)}, \dots, b_0^{(o_{n_o})}\} = \mathcal{S}_{b_0},
\end{equation} 
and each designated $b_0$ vertex is connected to: 
\begin{equation}
    \label{eq:C_3}
    N_{b_0^{(o_i)}} = \mathcal{\bar S}_{b_0}^{(o_i)} \cup \Gamma^{(o_{n_o})}, \; \; \forall b_0^{(o_i)} \in \mathcal{S}_{b_0}.
\end{equation}
    
Hence, by performing a Pauli-Z measurement on each of the vertices $\gamma \in \Gamma^{(o_{n_o})}$, the neighborhood of each designated $b_0^{(o_i)}$ vertex is updated as follows:
\begin{equation}
    \label{eq:C_4}
    N_{b_0^{(o_i)}}' = \mathcal{\bar S}_{b_0}^{(o_i)} , \; \; \forall b_0^{(o_i)} \in \mathcal{S}_{b_0}.
\end{equation}
    
The proof follows by noticing that the resulting graph is a collection of $n_o$ disconnected star graph states, each one centered in a $b_0^{(o_i)}$. 
Accordingly, being LU equivalent to a GHZ state, each star graph state allows the extraction of a single Bell state between the center $b_0^{(o_i)}$ and one of its adjacent vertices in $\mathcal{\bar S}_{b_0}^{(o_i)}$ by performing Pauli-Z measurements on the remaining $\hat{\kappa}_b - 2$ leaves ($\hat{\kappa}_b - 2$ measurements per star, $n_o(\hat{\kappa}_b-2)$ in total).
    
To show that $n_o$ is also the theoretical maximum number of instantiable Bell states in the operating regime of $\hat{\kappa}_b\geq 2$, we refer to the Schmidt Measure $\mathbf{P}(\ket{G})$ of a graph state $\ket{G}$ \cite{EisBriHan-01}. Specifically, for a two-colorable graph state $\ket{G}$ with associated graph $G=(V,E)$ and vertex set $V = V_o \cup V_c$ the following bounds hold \cite{HeiEisBri-04}:
\begin{align}
    \label{eq:C_5}
    \dfrac{1}{2}\text{rank}{(\Gamma_G)} \leq \mathbf{P}(\ket{G}) \leq \min\{|V_o|, |V_c|\},
\end{align}
where $\text{rank}{(\Gamma_G)}$ represents the rank over the binary field GF(2) of the adjacency matrix of the graph $G$ and $\mathbf{P}(\ket{G})$ is the Schmidt Measure of the graph state $\ket{G}$. Accordingly, the Schmidt measure has a closed form for the initial GTL resource state.
Since operations on the individual qubits, as in our procedure even when a node holds several of them, cannot extract more than $\mathbf{P}(\ket{G}) \leq |V_o| = n_o$ Bell states, the number of extractable Bell states is also the maximum.

\section{Proof of Proposition~\ref{prop:noise_map_rolling}}

\label{sec:noise_map_rolling}
Let us consider the ordered Pauli-X measurement sequence $\mathcal{O}_x =  \{\mathbf{M}_{x,\pm}^{(o_1)}, \mathbf{M}_{x,\pm}^{(o_2)}, \dots \mathbf{M}_{x,\pm}^{(o_{n_o})} \}$ acting on the orchestration qubits of the  GTL resource state. We analyze a generic peer qubit $c_j\in V_c$ adjacent to an orchestration qubit $o_i\in V_o$ and derive the corresponding NSF-updated depolarizing noise map after commuting local depolarizing noise through ${\mathcal{O}}_x$.
We recall that, the set of rolled vertices $\Gamma^{(o_m)}$, $m \in \{1, \dots, n_o\}$, can be updated at each measurement step, thus the support vertex $b_0^{(o_m)}$ can be chosen as left or right bridge accordingly, the left neighbors of $o_m$ being the rolled set $\Gamma^{(o_{m-1})}$. Hence, the set of non-$b_0$ vertices at measurement stage $m$ can be expressed according to Eq.~\eqref{eq:non_b0_bridges}.

When an orchestration qubit $o_{i-1}$ (if any) is measured in the X basis, the map of qubit $c_j$ is updated according to the neighborhood of the measured vertex. According to Tab.~\ref{tab:nsf_update_rules}, there are two possible update rules for the depolarizing map on $c_j$:
\begin{equation}
    \label{eq:P_1}
    \mathbf{M}_{x,\pm}^{(o_{i-1})} \rightarrow \mathbf{\tilde{\Lambda}}_{c_j,\alpha\beta} = \begin{cases}
        Z_{c_j}^{\beta} \prod_{k \in N'_{c_j}} Z_k^{\alpha} & \text{if } c_j = b_0^{(o_{i-1})} \\
        Z_{c_j}^{\alpha} \prod_{k \in N'_{c_j}} Z_k^{\beta} & \text{otherwise}
    \end{cases}.
\end{equation}
For any previous measurement step whose measured vertex is not adjacent to $c_j$, the map is unchanged.
When qubit $o_{i-1}$ is measured, the depolarizing noise map is then updated as:
\begin{align}
    \label{eq:P_2}
    &\mathcal{\tilde D}^{(o_{i-1})}_{c_j}(\varrho) = p\varrho + \dfrac{(1-p)}{4} \cdot \\
    & \begin{cases} 
    \underset{\alpha,\beta \in \{0,1\}} {\sum} \bigl(  Z_{c_j}^{\beta} \underset{k \in \tilde N}{\prod} Z_{k}^{\alpha} \bigr)
    \varrho 
    \bigl(  Z_{c_j}^{\beta} \underset{k \in \tilde N}{\prod} Z_{k}^{\alpha} \bigr)
    & \text{if } c_j = b_0^{(o_{i-1})} 
    \\
    \underset{\alpha,\beta \in \{0,1\}} {\sum} \bigl(  Z_{c_j}^{\alpha} \underset{k \in \tilde N}{\prod} Z_{k}^{\beta} \bigr)
    \varrho 
    \bigl(  Z_{c_j}^{\alpha} \underset{k \in \tilde N}{\prod} Z_{k}^{\beta} \bigr)
    & \text{otherwise} 
    \end{cases} \nonumber ,
\end{align}
where the post-measurement neighborhood $\tilde N$ is given accordingly to Th.~\ref{th:entanglement_rolling_effect}:
\begin{equation}
    \label{eq:P_3}
    \tilde N = \begin{cases}
        \Gamma^{(o_{i-1})} \cup \mathcal{\bar S}_{b_0}^{(o_{i-1})} & \text{if } c_j = b_0^{(o_{i-1})}  \\
        \{o_i\} \cup \bigcup_{\ell=1}^{{i-1}} \{b_0^{(o_\ell)}\} & \text{if } c_j \in \Gamma^{(o_{i-1})} \\
        \{b_0^{(o_{i-1})}\} & \text{if } c_j \in \mathcal{\bar S}_{b_0}^{(o_{i-1})}
    \end{cases}.
\end{equation}
When the considered $o_i$ qubit is measured, the noise operator updates according to the same rules above. Notably, if $c_j \notin \Gamma^{(o_m)}$, $m\in\{i, \dots, n_o\}$, its map is no longer updated, according to the resulting post-measurement neighborhood, except for the support vertices, which remain adjacent to the current rolled set. Consequently, the depolarizing map is updated as follows:
\begin{align}
    \label{eq:P_4}
    &\mathcal{\tilde D}^{(o_{i})}_{c_j}(\varrho) = p\varrho + \dfrac{(1-p)}{4} \cdot \\
    & \begin{cases} 
    \underset{\alpha,\beta \in \{0,1\}} {\sum} \bigl(  Z_{c_j}^{\beta} \underset{k \in \tilde N}{\prod} Z_{k}^{\alpha} \bigr)
    \varrho 
    \bigl(  Z_{c_j}^{\beta} \underset{k \in \tilde N}{\prod} Z_{k}^{\alpha} \bigr)
    & \text{if } c_j = b_0^{(o_{i})} 
    \\
    \underset{\alpha,\beta \in \{0,1\}} {\sum} \bigl(  Z_{c_j}^{\alpha} \underset{k \in \tilde N}{\prod} Z_{k}^{\beta} \bigr)
    \varrho 
    \bigl(  Z_{c_j}^{\alpha} \underset{k \in \tilde N}{\prod} Z_{k}^{\beta} \bigr)
    & \text{otherwise} 
    \end{cases} \nonumber ,
\end{align}
where the post-measurement neighborhood $\tilde N$ is updated accordingly:
\begin{equation}
    \label{eq:P_5}
    \tilde N = 
    \begin{cases}
        \Gamma^{(o_i)} \cup \mathcal{\bar S}_{b_0}^{(o_m)} & \begin{aligned} \text{if } &c_j = b_0^{(o_m)} \, , \\ &m\in\{i-1, i\} \end{aligned} \\
        \{o_{i+1}\} \cup \underset{\ell=1}{\bigcup^{i}} \{b_0^{(o_\ell)}\} & \text{if } c_j \in \Gamma^{(o_i)} \\
        \{b_0^{(o_m)}\} & \begin{aligned} \text{if } &c_j \in \mathcal{\bar S}_{b_0}^{(o_m)}, \\ &m \in\{i-1, i\} \end{aligned}
    \end{cases}.
\end{equation}

By applying the same update rules until the last measurement, and by noting that $\{\beta,\alpha\}$ can be relabeled as $\{\alpha,\beta\}$ without loss of generality, the overall noise map on peer qubit $c_j$ is given by:
\begin{align}
    \label{eq:P_6}
    \mathcal{\tilde D}&_{c_j}(\varrho) = p\varrho + \dfrac{(1-p)}{4} \cdot \\
    &\sum_{\alpha,\beta \in \{0,1\}} \bigl(  Z_{c_j}^{\alpha} \underset{k \in \tilde N}{\prod} Z_{k}^{\beta} \bigr) \varrho \bigl(  Z_{c_j}^{\alpha} \underset{k \in \tilde N}{\prod} Z_{k}^{\beta} \bigr) \nonumber 
\end{align}
where the post-measurement neighborhood $\tilde N$ is given by:
\begin{equation}
    \label{eq:P_7}
    \tilde N = \begin{cases}
        \Gamma^{(o_{n_o})} \cup \mathcal{\bar S}_{b_0}^{(o_m)} & \begin{aligned} \text{if } &c_j = b_0^{(o_m)} \, , \\ &m\in\{1, \dots, n_o\} \end{aligned} \\
        \underset{\ell=1}{\bigcup^{n_o}} \
{b_0^{(o_\ell)}} & \text{if } c_j \in \Gamma^{(o_{n_o})} \\
        \{b_0^{(o_m)}\} & \begin{aligned} \text{if } &c_j \in \mathcal{\bar S}_{b_0}^{(o_m)}, \\ &m \in\{1, \dots, n_o\} \end{aligned}
    \end{cases}.
\end{equation}

By expanding the coefficients, the full expression of the final noise map is as follows:
\begin{align}
    \label{eq:P_8}
    \mathcal{\tilde D}_{c_j}(\varrho)=p\varrho &+ \tfrac{(1-p)}{4} \bigl[ \varrho +  Z_{c_j} \varrho Z_{c_j} + \bigl(\prod_{k \in \tilde N} Z_k \bigr)  \varrho \bigl(\prod_{k \in \tilde N} Z_k \bigr) + \nonumber \\ 
    &+ \bigl( Z_{c_j} \prod_{k \in \tilde N} Z_k \bigr) \varrho \bigl( Z_{c_j} \prod_{k \in \tilde N} Z_k \bigr) \bigr].
\end{align}

The same approach can be carried on for the depolarizing noise map of each orchestration qubit $o_i \in V_o$. If $o_{i-1}$ is measured, its noise map is updated according to Tab.~\ref{tab:nsf_update_rules} as follows:
\begin{align}
    \label{eq:P_9}
    \mathcal{\tilde D}^{(o_{i-1})}_{o_i}(\varrho) &= p\varrho + \dfrac{(1-p)}{4} \cdot \\ 
    &\sum_{\alpha,\beta \in \{0,1\}} \bigl(  Z_{o_i}^{\alpha} \underset{k \in \tilde N}{\prod} Z_{k}^{\beta} \bigr) \varrho \bigl(  Z_{o_i}^{\alpha} \underset{k \in \tilde N}{\prod} Z_{k}^{\beta} \bigr), \nonumber
\end{align}
where, regardless of the arbitrary choice of $\Gamma^{(o_{i-1})}$ as left or right bridge set, the post-measurement neighborhood $\tilde N$ is given by:
\begin{equation}
    \tilde N = \mathcal{R}_B^{(o_i)} \cup \Gamma^{(o_{i-1})}.
\end{equation}

When $o_i$ is measured, its noise map is then updated as:
\begin{align}
    \label{eq:P_10}
    \mathcal{\tilde D}^{(o_{i})}_{o_i}&(\varrho) = p\varrho + \dfrac{(1-p)}{4} \cdot \\
    &\sum_{\alpha,\beta \in \{0,1\}} \bigl(  Z_{{b_0}^{(o_i)}}^{\alpha} \underset{\substack{k\in N_{b_0}^{(o_i)}\\k\ne o_i}}{\prod} Z_{k}^{\alpha} \bigr) \varrho \bigl(  Z_{{b_0}^{(o_i)}}^{\alpha} \underset{\substack{k\in N_{b_0}^{(o_i)}\\k\ne o_i}}{\prod} Z_{k}^{\alpha} \bigr), \nonumber
\end{align}
where the pre-measurement neighborhood $N_{b_0}^{(o_i)}$ is:
\begin{equation}
    \label{eq:P_11}
    N_{b_0}^{(o_i)} = 
    \begin{cases}
        \{o_i\} \cup \{o_{i+1}\} & \text{if} \; b_0^{(o_i)} \in \mathcal{R}_B^{(o_i)} \;\; (o_i \neq o_{n_o})\\
        \{o_i\} & \text{if} \; b_0^{(o_i)} \in \mathcal{R}_B^{(o_i)} \;\; (o_i = o_{n_o})\\
        \bigcup_{\ell=1}^{i-1}\{b_0^{(o_\ell)}\} \cup \{o_{i}\} & \text{if} \; b_0^{(o_i)} \in \Gamma^{(o_{i-1})}
    \end{cases}.
\end{equation}
For a right support at step $i$, Eq.~\eqref{eq:P_11} leaves a phase flip on that support and, if present, on $o_{i+1}$. Each subsequent right step transfers the latter to its own support and the next orchestration qubit. An all-right suffix therefore leaves $\prod_{\ell=i}^{n_o}Z_{b_0^{(o_\ell)}}^\alpha$. Conversely, if $b_0^{(o_i)} \in \Gamma^{(o_{i-1})}$, Eq.~\eqref{eq:P_11} gives $N_{b_0}^{(o_i)} \setminus \{o_i\} = \bigcup_{\ell=1}^{i-1}\{b_0^{(o_\ell)}\}$, so the residual operator $Z_{b_0^{(o_i)}}^\alpha \prod_{\ell=1}^{i-1} Z_{b_0^{(o_\ell)}}^\alpha$ acts only on support vertices, which are neither measured nor adjacent to an orchestration qubit afterwards (Eq.~\eqref{eq:P_5}), and it is no longer updated. Hence, if $j \geq i$ is the first step such that $b_0^{(o_j)} \in \Gamma^{(o_{j-1})}$, the operator $Z_{b_0^{(o_i)}}^\alpha \cdots Z_{b_0^{(o_{j-1})}}^\alpha Z_{o_j}^\alpha$ obtained after the measurements of $o_i, \dots, o_{j-1}$ becomes:
\begin{equation*}
    Z_{b_0^{(o_i)}}^\alpha \cdots Z_{b_0^{(o_{j-1})}}^\alpha \, Z_{b_0^{(o_j)}}^\alpha \prod_{\ell=1}^{j-1} Z_{b_0^{(o_\ell)}}^\alpha = Z_{b_0^{(o_j)}}^\alpha \prod_{\ell=1}^{i-1} Z_{b_0^{(o_\ell)}}^\alpha,
\end{equation*}
since $Z^2 = \mathbb{I}$. The final noise map is thus given by:
\begin{align}
    \label{eq:P_12}
    \mathcal{\tilde D}_{o_i}(\varrho) &= p\varrho + \dfrac{(1-p)}{2} \cdot \\
    &\Bigl[
        \varrho + \bigl( \prod_{k \in \bar N} Z_k \bigr) \varrho \bigl( \prod_{k \in \bar N} Z_k \bigr)
    \Bigr], \nonumber
\end{align}
with $\bar N$ the set of Prop.~\ref{prop:noise_map_rolling}: $\bar N = \bigcup_{\ell=i}^{n_o}\{b_0^{(o_\ell)}\}$ if no left support is selected from step $i$ on, and $\bar N = \{b_0^{(o_j)}\} \cup \bigcup_{\ell=1}^{i-1}\{b_0^{(o_\ell)}\}$ otherwise. This concludes the proof.

\section{Proof of Corollary~\ref{cor:bell_fidelity}}
\label{sec:bell_fidelity}
Let $b=b_0^{(o_m)}$, let $a$ be the peer of $\mathcal{\bar S}_{b_0}^{(o_m)}$ not measured in Stage~2, and let $K_a=X_aZ_b$ and $K_b=Z_aX_b$ be the stabilizer operators of the graph state $\ket{\Phi}$ of the edge $\{a,b\}$ (Def.~\ref{def:stabilizer_operators}). Since $\ket{\Phi}\bra{\Phi}=\frac{1}{4}(\mathbb{I}+K_a+K_b+K_aK_b)$~\cite{HeiDurEis-06}, the fidelity of the pair is
\begin{equation}
    \label{eq:D_1}
    F_m = \tfrac{1}{4}\bigl(1+\langle K_a\rangle+\langle K_b\rangle+\langle K_aK_b\rangle\bigr),
\end{equation}
where $\langle K\rangle$ is the expectation value of $K$ on the output state $\tilde{\mathcal{D}}_1\cdots\tilde{\mathcal{D}}_n\varrho'$ of Eq.~\eqref{eq:noisy_stabilizer_formalism_updated}, $\varrho'$ being the noiseless output of the two-stage resolution, for which $\langle K\rangle=1$, since $\{a,b\}$ is an isolated edge of the output graph (Cor.~\ref{cor:maximal_ent_extraction}).

By Prop.~\ref{prop:noise_map_rolling} and the Remark in Sec.~\ref{sec:06-A}, each updated map reads
\begin{equation}
    \label{eq:D_2}
    \tilde{\mathcal{D}}_j(\varrho) = p\varrho+\dfrac{1-p}{4}\sum_{\alpha,\beta\in\{0,1\}}\mathbf{\tilde{\Lambda}}_{j,\alpha\beta}\,\varrho\,\mathbf{\tilde{\Lambda}}_{j,\alpha\beta},
\end{equation}
with $\mathbf{\tilde{\Lambda}}_{j,\alpha\beta}=\mathbf{\tilde{\Lambda}}_{j,10}^{\alpha}\mathbf{\tilde{\Lambda}}_{j,01}^{\beta}$ products of Pauli-Z operators, and $\mathbf{\tilde{\Lambda}}_{j,10}$ or $\mathbf{\tilde{\Lambda}}_{j,01}$ equal to $\mathbb{I}$ for the measured qubits. Each noise operator either commutes or anticommutes with $K$, and $\mathbf{\tilde{\Lambda}}_{j,11}$ anticommutes with $K$ if and only if exactly one of $\mathbf{\tilde{\Lambda}}_{j,10}$ and $\mathbf{\tilde{\Lambda}}_{j,01}$ does. Hence, either none or exactly two of the three noise operators anticommute with $K$. In the first case, the map leaves $\langle K\rangle$ unchanged; in the second, it multiplies $\langle K\rangle$ by $p+\frac{1-p}{4}(1+1-1-1)=p$. Since $\langle K\rangle=1$ without noise, it follows that
\begin{equation}
    \label{eq:D_3}
    \langle K\rangle = p^{\,n_K},
\end{equation}
where $n_K$ is the number of maps with a noise operator anticommuting with $K$.

It suffices to restrict the noise operators to $\{a,b\}$. Each endpoint map contains $Z_a$, $Z_b$ and $Z_aZ_b$ and therefore contributes to all three counts. The remaining maps acting on the pair contribute through $Z_b$ alone. They come from the $\hat{\kappa}_b-2$ measured peers in the group of $b$, the $\hat{\kappa}_b$ peers of the final rolled set $\Gamma^{(o_{n_o})}$, and the $n_b$ orchestration qubits $o_i$ whose set $\bar N$ contains $b$ (Prop.~\ref{prop:noise_map_rolling}). Since $Z_b$ commutes with $K_a$ and anticommutes with $K_b$ and $K_aK_b$, Eq.~\eqref{eq:D_3} yields
\begin{equation}
    \label{eq:D_4}
    \langle K_a\rangle = p^{2}, \qquad \langle K_b\rangle = \langle K_aK_b\rangle = p^{\,2\hat{\kappa}_b+n_b}.
\end{equation}

The reduced state is diagonal in the graph-state basis, which is locally equivalent to the Bell basis. Since all three stabilizer expectations in Eq.~\eqref{eq:D_4} are nonnegative, the target state has the largest weight $F_m$, so the pair is entangled and distillable if and only if $F_m>1/2$.

To identify $n_b$, fix $o_i$ in the initial graph $G$. Removing $o_i$ separates the peer groups to its left from those to its right. Before step $i$, all earlier supports and the current rolled set lie to its left (Th.~\ref{th:entanglement_rolling_effect} and the Remark following it). If every step from $i$ onward takes a right support, the rolled set stays on the left, and $\bar N$ consists of the supports from steps $i,\ldots,n_o$, all on the right. Otherwise let $j\ge i$ be the first step at or after $i$ with a left support. Step $j$ takes its support from the rolled set on the left and moves that set to $\mathcal R_B^{(o_j)}$ on the right. The intervening right supports and all subsequent supports lie on the right. Thus the supports on the left are exactly those from steps $1,\ldots,i-1$ and step $j$, which form $\bar N$ in the second case of Prop.~\ref{prop:noise_map_rolling}.

In both cases, $b\in\bar N$ exactly when $o_i$ separates $b$ from any $\gamma\in\Gamma^{(o_{n_o})}$ in $G$. A shortest path from $b$ to $\gamma$ traverses the intervening peer groups in order and hence visits exactly these orchestration qubits. Such a path alternates between orchestration qubits and peers, with a bridge between each two successive orchestration qubits. Hence $n_b$ is one plus the number of interior bridges, namely $\pi(b,\gamma)$ by Def.~\ref{def:peer_proximity}. Substituting in Eqs.~\eqref{eq:D_4} and~\eqref{eq:D_1} proves Eq.~\eqref{eq:bell_fidelity}.

\end{appendices}

\bibliography{biblio}

@article{PeaMazCal-26-journal,
  title={An extensible quantum network simulator built on ns-3: Q2NS design and evaluation},
  author={Pearson, Adam and Mazza, Francesco and Caleffi, Marcello and Cacciapuoti, Angela Sara},
  journal={Computer Networks},
  volume={283},
  pages={112292},
  doi={10.1016/j.comnet.2026.112292},
  year={2026},
  publisher={Elsevier}
}

@article{CacCal-26,
  title={A quantum internet protocol suite beyond layering},
  author={Cacciapuoti, Angela Sara and Caleffi, Marcello},
  journal={IEEE TNSE},
  volume={13},
  pages={9170--9187},
  year={2026},
  publisher={IEEE},
  doi={10.1109/TNSE.2026.3679795},
  note = {invited paper}
}

@article{CalCac-25,
  title={{Quantum Internet Architecture: unlocking Quantum-Native Routing via Quantum Addressing}},
  author={Caleffi, Marcello and Cacciapuoti, Angela Sara},
  journal={IEEE TCOM},
  volume={74},
  pages={3577--3599},
  doi = {10.1109/TCOMM.2025.3650397},
  year={2026},
  note = {invited paper}
}

@misc{CheCacCal-26,
      title={Quantum Routing Beyond Pathfinding: Multipartite Entanglement Complementation}, 
      author={Si-Yi Chen and Angela Sara Cacciapuoti and Marcello Caleffi},
      year={2026},
      eprint={2604.13834},
      archivePrefix={arXiv},
      primaryClass={quant-ph},
      url={https://arxiv.org/abs/2604.13834}, 
}

@inproceedings{MazZhaChu-25,
      title={{Simulation of Entanglement-Enabled Connectivity in QLANs using SeQUeNCe}}, 
      author={Francesco Mazza and Caitao Zhan and Joaquin Chung and Rajkumar Kettimuthu and Marcello Caleffi and Angela Sara Cacciapuoti},
      year={2025},
      pages={1--6},
      doi={10.1109/ICC52391.2025.11434940},
      booktitle = {{IEEE ICC 2025}},
}

@article{CirZolKim-97,
	title={Quantum state transfer and entanglement distribution among distant nodes in a quantum network},
	author={Cirac, Juan Ignacio and Zoller, Peter and Kimble, H Jeff and others},
	journal = {Phys. Rev. Lett.},
	volume={78},
	number={16},
	pages={3221},
	year={1997},
	publisher={APS}
}

@article{EisBriHan-01,
   title={Schmidt measure as a tool for quantifying multiparticle entanglement},
   volume={64},
   ISSN={1094-1622},
   url={http://dx.doi.org/10.1103/PhysRevA.64.022306},
   DOI={10.1103/physreva.64.022306},
   number={2},
   pages={022306},
   journal={Phys. Rev. A},
   publisher={American Physical Society (APS)},
   author={Eisert, Jens and Briegel, Hans-J{\"u}rgen},
   year={2001},
   month=jul }

@article{JozLin-03,
    author = {Jozsa, Richard  and Linden, Noah },
    title = {On the role of entanglement in quantum-computational speed-up},
    journal = {Proc. R. Soc. Lond. A},
    volume = {459},
    number = {2036},
    pages = {2011-2032},
    year = {2003},
    doi = {10.1098/rspa.2002.1097},
}

@article{HeiEisBri-04,
  title={Multiparty entanglement in graph states},
  author={Hein, Marc and Eisert, Jens and Briegel, Hans J},
  journal={Phys. Rev. A},
  volume={69},
  number={6},
  pages={062311},
  year={2004},
  publisher={APS}
}

@article{PirDur-19,
	year = 2019,
	month = {mar},
	publisher = {{IOP} Publishing},
	volume = {21},
	number = {3},
	pages = {033003},
	author = {A Pirker and W Dür},
	title = {A quantum network stack and protocols for reliable entanglement-based networks},
	journal = {New Journal of Physics},
}

@article{CacCalVan-20,
 title={When entanglement meets classical communications: Quantum teleportation for the quantum internet},
	author={Cacciapuoti, Angela Sara and Caleffi, Marcello and Van Meter, Rodney and Hanzo, Lajos},
	journal={IEEE TCOM},
	volume={68},
	number={6},
	pages={3808--3833},
	year={2020},
	publisher={IEEE},
	note = {invited paper},
}

@article{RamPirDur-21,
	author = {Miguel-Ramiro, J. and Pirker, A. and D{\"u}r, W.},
	year = {2021},
	title = {Genuine quantum networks with superposed tasks and addressing},
	journal = {npj Quantum Information},
	volume = {7},
	pages = {135},
	doi = {10.1038/s41534-021-00472-5}
}

@article{IllCalMan-22,
    title={Quantum Internet Protocol Stack: a Comprehensive Survey},
    author={Illiano, Jessica and Caleffi, Marcello and Manzalini, Antonio and Cacciapuoti, Angela Sara},
    journal={Computer Networks},
    volume={213},
    pages={109092},
    doi={10.1016/j.comnet.2022.109092},
    year={2022}
}

@article{HeiDurEis-06,
  title={Entanglement in graph states and its applications},
  author={Hein, Marc and D{\"u}r, Wolfgang and Eisert, Jens and Raussendorf, Robert and {Van den Nest}, Maarten and Briegel, Hans-J{\"u}rgen},
  journal={arXiv preprint quant-ph/0602096},
  year={2006}
}

@ARTICLE{IllCalVis-23,
    author={Illiano, Jessica and Caleffi, Marcello and others},
    journal={IEEE TCOM}, 
    title={{Quantum MAC}: Genuine Entanglement Access Control via Many-Body Dicke States}, 
    year={2024},
    volume={72},
    number={4},
    pages={2090--2105},
    doi={10.1109/TCOMM.2023.3344140}
}

@article{AviRozWeh-23,
  title = {Analysis of multipartite entanglement distribution using a central quantum-network node},
  author = {Avis, Guus and Rozp\ifmmode \mbox{\k{e}}\else \k{e}\fi{}dek, Filip and Wehner, Stephanie},
  journal = {Phys. Rev. A},
  volume = {107},
  issue = {1},
  pages = {012609},
  numpages = {36},
  year = {2023},
  month = {Jan},
  publisher = {American Physical Society},
  doi = {10.1103/PhysRevA.107.012609},
  url = {https://link.aps.org/doi/10.1103/PhysRevA.107.012609}
}

@ARTICLE{CacIllCal-23,
  author={Cacciapuoti, Angela Sara and Illiano, Jessica and Caleffi, Marcello},
  journal={IEEE Network}, 
  title={Quantum Internet Addressing}, 
  year={2024},
  volume={38},
  number={1},
  pages={104--111},
  doi={10.1109/MNET.2023.3328393}
}

@INPROCEEDINGS{ChuRamAni-24,
  author={Chung, Joaquin and others},
  booktitle={OFC}, 
  title={{Orchestration of Entanglement Distribution over a Q-LAN using the IEQNET Controller}}, 
  year={2024},
  pages={M3Z.4},
  doi={10.1364/OFC.2024.M3Z.4},
}

@article{CirEkeHue-99,
    author = {Cirac, J. I. and Ekert, A. K. and Huelga, S. F. and others},
    doi = {10.1103/PhysRevA.59.4249},
    issue = {6},
    journal = {Phys. Rev. A},
    month = {Jun},
    numpages = {0},
    pages = {4249--4254},
    publisher = {American Physical Society},
    title = {Distributed quantum computation over noisy channels},
    url = {http://link.aps.org/doi/10.1103/PhysRevA.59.4249},
    volume = {59},
    year = {1999}
}

@article{HayMor-15,
  title = {Verifiable Measurement-Only Blind Quantum Computing with Stabilizer Testing},
  author = {Hayashi, Masahito and Morimae, Tomoyuki},
  journal = {Phys. Rev. Lett.},
  volume = {115},
  issue = {22},
  pages = {220502},
  numpages = {5},
  year = {2015},
  month = {Nov},
  publisher = {American Physical Society},
  doi = {10.1103/PhysRevLett.115.220502},
  url = {https://link.aps.org/doi/10.1103/PhysRevLett.115.220502}
}

@article{GisRibTit-02,
  title = {Quantum cryptography},
  author = {Gisin, Nicolas and Ribordy, Gr\'egoire and Tittel, Wolfgang and others},
  journal = {Rev. Mod. Phys.},
  volume = {74},
  issue = {1},
  pages = {145--195},
  numpages = {0},
  year = {2002},
  month = {Mar},
  publisher = {American Physical Society},
  doi = {10.1103/RevModPhys.74.145},
  url = {https://link.aps.org/doi/10.1103/RevModPhys.74.145}
}

@article{PirAndBan-20, 
  title={Advances in quantum cryptography}, 
  volume={12}, 
  ISSN={1943-8206}, 
  url={http://dx.doi.org/10.1364/AOP.361502}, 
  DOI={10.1364/aop.361502}, 
  number={4}, 
  journal={Advances in Optics and Photonics}, publisher={Optica Publishing Group}, 
  author={Pirandola, S. and Andersen, U. L. and Banchi, L. and others}, 
  year={2020}, 
  month=dec, 
  pages={1012} 
}

@article{GioLloMac-11, 
  title={Advances in quantum metrology}, 
  volume={5}, 
  ISSN={1749-4893}, 
  url={http://dx.doi.org/10.1038/nphoton.2011.35}, 
  DOI={10.1038/nphoton.2011.35}, 
  number={4}, 
  journal={Nature Photonics}, 
  publisher={Springer Science and Business Media LLC}, 
  author={Giovannetti, Vittorio and Lloyd, Seth and Maccone, Lorenzo}, 
  year={2011}, 
  month=mar, 
  pages={222--229}
}

@article{KesLovSus-14,
    title = {Quantum Error Correction for Metrology},
    author = {Kessler, E. M. and Lovchinsky, I. and Sushkov, A. O. and others},
    journal = {Phys. Rev. Lett.},
    volume = {112},
    issue = {15},
    pages = {150802},
    numpages = {5},
    year = {2014},
    month = {Apr},
    publisher = {American Physical Society},
    doi = {10.1103/PhysRevLett.112.150802},
    url = {https://link.aps.org/doi/10.1103/PhysRevLett.112.150802}
}

@article{SekWolDur-20,
    title = {Optimal distributed sensing in noisy environments},
    author = {Sekatski, P. and W\"olk, S. and D\"ur, W.},
    journal = {Phys. Rev. Research},
    volume = {2},
    issue = {2},
    pages = {023052},
    numpages = {8},
    year = {2020},
    month = {Apr},
    publisher = {American Physical Society},
    doi = {10.1103/PhysRevResearch.2.023052},
    url = {https://link.aps.org/doi/10.1103/PhysRevResearch.2.023052}
}

@article{AzuEcoElk-23,
  title = {Quantum repeaters: From quantum networks to the quantum internet},
  author = {Azuma, Koji and Economou, Sophia E. and Elkouss, David and others},
  journal = {Rev. Mod. Phys.},
  volume = {95},
  issue = {4},
  pages = {045006},
  numpages = {66},
  year = {2023},
  month = {Dec},
  publisher = {American Physical Society},
  doi = {10.1103/RevModPhys.95.045006},
  url = {https://link.aps.org/doi/10.1103/RevModPhys.95.045006}
}

@ARTICLE{GiaWinCon-25,
  author={Giani, Andrea and Win, Moe Z. and Conti, Andrea},
  journal={IEEE JSAIT}, 
  title={Quantum Sensing and Communication via Non-Gaussian States}, 
  year={2025},
  volume={6},
  number={},
  pages={18-33},
  doi={10.1109/JSAIT.2024.3491692}
}

@techreport{CacCalIll-26,
    number =    {draft-cacciapuoti-qirg-quantum-native-architecture-02},
    type =      {Internet-Draft},
    institution =   {Internet Engineering Task Force},
    publisher = {Internet Engineering Task Force},
    note =      {Work in Progress},
    url =       {https://datatracker.ietf.org/doc/draft-cacciapuoti-qirg-quantum-native-architecture/02/},
    author =    {Angela Sara Cacciapuoti and Marcello Caleffi and Jessica Illiano and C. De Risi and A. Abane and Joaquin Chung},
    title =     {{Quantum-Native Architectural Tenets and Philosophy for the Quantum Internet}},
    pagetotal = 27,
    year =      2026,
    month =     oct,
    day =       17,
}

@article{AigRuiDur-25,
  title = {Qudit noisy stabilizer formalism},
  author = {Aigner, Paul and Mor-Ruiz, Maria Flors and D\"ur, Wolfgang},
  journal = {Phys. Rev. A},
  volume = {112},
  issue = {2},
  pages = {022402},
  numpages = {23},
  year = {2025},
  month = {Aug},
  publisher = {American Physical Society},
  doi = {10.1103/gqfw-x72s},
  url = {https://link.aps.org/doi/10.1103/gqfw-x72s}
}

@article{MazCalCac-25,
  title={Intra-{QLAN} connectivity via graph states: Beyond the physical topology},
  author={Mazza, Francesco and Caleffi, Marcello and Cacciapuoti, Angela Sara},
  journal={IEEE TNSE},
  volume={12},
  number={2},
  pages={870--887},
  doi={10.1109/TNSE.2024.3520856},
  year={2025},
  publisher={IEEE}
}

@article{RuiDur-23,
  title={Noisy stabilizer formalism},
  author={Mor-Ruiz, Maria Flors and D{\"u}r, Wolfgang},
  journal={Phys. Rev. A},
  volume={107},
  number={3},
  pages={032424},
  year={2023},
  publisher={APS}
}

@article{CalAmoFer-22,
  title = {Distributed quantum computing: A survey},
journal = {Computer Networks},
volume = {254},
pages = {110672},
year = {2024},
issn = {1389-1286},
doi = {https://doi.org/10.1016/j.comnet.2024.110672},
url = {https://www.sciencedirect.com/science/article/pii/S1389128624005048},
author = {Marcello Caleffi and Michele Amoretti and Davide Ferrari and Jessica Illiano and Antonio Manzalini and Angela Sara Cacciapuoti},
}

@article{HahPapEis-19,
  title={Quantum network routing and local complementation},
  author={Hahn, Frederik and Pappa, A and Eisert, Jens},
  journal={npj Quantum Information},
  volume={5},
  number={1},
  pages={76},
  year={2019},
  publisher={Nature Publishing Group UK London}
}

@ARTICLE{AbaCubMai-25,
  author={Abane, Amar and others},
  journal={IEEE TQE}, 
  title={Entanglement Routing in Quantum Networks: A Comprehensive Survey}, 
  year={2025},
  volume={6},
  number={},
  pages={1-39},
  doi={10.1109/TQE.2025.3541123}
}

@article{IneVarSca-23,
  title={Optimal entanglement distribution policies in homogeneous repeater chains with cutoffs},
  author={I{\~n}esta, {\'A}lvaro G and others},
  journal={npj Quantum Information},
  volume={9},
  number={1},
  pages={46},
  year={2023},
  publisher={Nature Publishing Group UK London}
}

@article{Kim-08,
    year = {2008},
    pages = {1023--1030},
    title = {The quantum internet},
    author = {Kimble, H Jeff},
    number = {7198},
    volume = {453},
    journal = {Nature},
    publisher = {Nature Publishing Group}
}

@article{DurLamHeu-17,
    year = {2017},
    pages = {043001},
    title = {Towards a quantum internet},
    author = {D{\"u}r, Wolfgang and Lamprecht, Raphael and Heusler, Stefan},
    number = {4},
    volume = {38},
    journal = {European Journal of Physics},
    publisher = {IOP Publishing}
}

@article{DahWeh-18,
  title={Transforming graph states using single-qubit operations},
  author={Dahlberg, Axel and Wehner, Stephanie},
  journal={Philos. Trans. R. Soc. A},
  volume={376},
  number={2123},
  pages={20170325},
  year={2018},
  publisher={The Royal Society Publishing}
}

@article{DahHelWeh-20,
  title = {Transforming graph states to {Bell}-pairs is {NP}-{Complete}},
  author = {Dahlberg, Axel and Helsen, Jonas and Wehner, Stephanie},
  journal = {Quantum},
  volume = {4},
  pages = {348},
  year = {2020},
  doi = {10.22331/q-2020-10-22-348}
}

@article{DahHelWeh-20b,
  title={How to transform graph states using single-qubit operations: computational complexity and algorithms},
  author={Dahlberg, Axel and Helsen, Jonas and Wehner, Stephanie},
  journal={Quantum Science and Technology},
  volume={5},
  number={4},
  pages={045016},
  year={2020},
  doi={10.1088/2058-9565/aba763}
}

@article{ThoRusMor-22,
  title={Efficient generation of entangled multiphoton graph states from a single atom},
  author={Thomas, Philip and others},
  journal={Nature},
  volume={608},
  number={7924},
  pages={677--681},
  year={2022},
  publisher={Nature Publishing Group UK London}
}

@article{ThoRusRem-24,
    author = {Thomas, Philip and others},
    year = {2024},
    journal = {Nature},
    volume = {629},
    title = {Fusion of deterministically generated photonic graph states},
    url = {https://doi.org/10.1038/s41586-024-07357-5},
    pages = {567--572},
    doi = {10.1038/s41586-024-07357-5}
}

@article{ButBarEco-17,
  title = {Deterministic Generation of All-Photonic Quantum Repeaters from Solid-State Emitters},
  author = {Buterakos, Donovan and Barnes, Edwin and Economou, Sophia E.},
  journal = {Phys. Rev. X},
  volume = {{7}},
  issue = {{4}},
  pages = {{041023}},
  numpages = {{10}},
  year = {{2017}},
  month = {{Oct}},
  publisher = {{American Physical Society}},
  doi = {{10.1103/PhysRevX.7.041023}},
  url = {https://link.aps.org/doi/10.1103/PhysRevX.7.041023}
}

@article{MazRamIll-25,
  title={Flexible Qubit Allocation of Network Resource States},
  author={Mazza, Francesco and Miguel-Ramiro, Jorge and Illiano, Jessica and Pirker, Alexander and Caleffi, Marcello and Cacciapuoti, Angela Sara and D{\"u}r, Wolfgang},
  journal={npj Quantum Information, in press},
  note={arXiv:2510.15776},
  year={2026}
}

@article{RamIllMaz-26,
  title={QPing: a Quantum Ping Primitive for Quantum Networks},
  author={Miguel-Ramiro, Jorge and Illiano, Jessica and Mazza, Francesco and Pirker, Alexander and Freund, Julia and Cacciapuoti, Angela Sara and Caleffi, Marcello and D{\"u}r, Wolfgang},
  journal={IEEE JSAC},
  volume={44},
  pages={4997--5011},
  doi={10.1109/JSAC.2026.3693981},
  year={2026}
}

@article{RuiWalDur-25,
  title={Imperfect quantum networks with tailored resource states},
  author={Mor-Ruiz, Maria Flors and Walln{\"o}fer, Julius and D{\"u}r, Wolfgang},
  journal={Quantum},
  volume={9},
  pages={1605},
  year={2025},
  publisher={Verein zur F{\"o}rderung des Open Access Publizierens in den Quantenwissenschaften}
}

@article{RuiRamWal-26,
  title={Merging-based quantum repeater},
  author={Mor-Ruiz, Maria Flors and Miguel-Ramiro, Jorge and Walln{\"o}fer, Julius and Coopmans, Tim and D{\"u}r, Wolfgang},
  journal={npj Quantum Information},
  volume={12},
  pages={197},
  year={2026},
  doi={10.1038/s41534-026-01340-w}
}

@article{RusBarEco-19,
  title={Generation of arbitrary all-photonic graph states from quantum emitters},
  author={Russo, Antonio and Barnes, Edwin and Economou, Sophia E},
  journal={New Journal of Physics},
  volume={21},
  number={5},
  pages={055002},
  year={2019},
  publisher={IOP Publishing}
}

@article{AzuTamLo-15,
  title={All-photonic quantum repeaters},
  author={Azuma, Koji and Tamaki, Kiyoshi and Lo, Hoi-Kwong},
  journal={Nature Communications},
  volume={6},
  number={1},
  pages={6787},
  year={2015},
  publisher={Nature Publishing Group UK London}
}

@article{BhaGoo-26,
  title={Distributing graph states with a photon-weaving quantum server},
  author={Bhatti, Daniel and Goodenough, Kenneth},
  journal={New Journal of Physics},
  volume={28},
  number={7},
  pages={074507},
  year={2026},
  doi={10.1088/1367-2630/ae83dc}
}

@article{MenFauCha-25,
  title={Temporal fusion of entangled resource states from a quantum emitter},
  author={Meng, Yijian and others},
  journal={Nature Communications},
  volume={16},
  number={1},
  pages={7602},
  year={2025},
  publisher={Nature Publishing Group UK London},
  doi={10.1038/s41467-025-62130-0}
}

@article{BenBraPop-96,
  title = {Purification of Noisy Entanglement and Faithful Teleportation via Noisy Channels},
  author = {Bennett, Charles H. and Brassard, Gilles and Popescu, Sandu and Schumacher, Benjamin and Smolin, John A. and Wootters, William K.},
  journal = {Phys. Rev. Lett.},
  volume = {76},
  number = {5},
  pages = {722--725},
  year = {1996},
  doi = {10.1103/PhysRevLett.76.722}
}

@misc{nsf,
  author       = "Walln{\"o}fer, Julius and others",
  title        = "{Noisy Graph States}",
  year         = 2025,
  howpublished = "Zenodo, \url{https://doi.org/10.5281/zenodo.10625616}",
}

@misc{repo,
  author       = "Mazza, Francesco and others",
  title        = "{gtl-rolling: Entanglement Rolling on generalized tree-like graph states}",
  year         = 2026,
  howpublished = "Zenodo, version v0.1.0, \url{https://doi.org/10.5281/zenodo.24554724}",
}

@article{LiYanBao-26,
    author = {Li, Yan and Yang, Yunfan and Bao, Zenghui and Yang, Jize and Sun, Luyan and Song, Yipu and Wu, Yukai and Zhang, Hongyi and Duan, Luming},
    title = {Deterministic and programmable fusion for the scalable generation of photonic graph states},
    journal = {Nature Physics},
    year = {2026},
    doi = {10.1038/s41567-026-03471-5},
    url = {https://doi.org/10.1038/s41567-026-03471-5}
}

@article{DurCir-00,
  title = {Classification of multiqubit mixed states: Separability and distillability properties},
  author = {D\"ur, W. and Cirac, J. I.},
  journal = {Phys. Rev. A},
  volume = {61},
  number = {4},
  pages = {042314},
  year = {2000},
  publisher = {American Physical Society},
  doi = {10.1103/PhysRevA.61.042314},
  url = {https://link.aps.org/doi/10.1103/PhysRevA.61.042314}
}

@article{IllDeRCalCac-26,
  title={Quantum Routers: A Switching-Fabric Framework for Quantum-Native Forwarding},
  author={Illiano, Jessica and De Risi, Caterina and Cacciapuoti, Angela Sara and Caleffi, Marcello},
  journal={arXiv preprint arXiv:2606.17773},
  year={2026}
}

\end{document}